\documentstyle[12pt]{article}
\scrollmode

\newtheorem{theorem}{Theorem}[section]

\newtheorem{lemma}{Lemma}[section]

\newtheorem{definition}{Definition}[section]

\def\R{I\!\! R}

\def\C{I\!\!\!\!C}
\def\H{I\!\! H}
\def\T{\hbox{\bf T}}
\def\r{I\!\! R}

\def\qed{\vbox to 5.8pt{\offinterlineskip\hrule
         \hbox to 5.8pt{\vrule height 5.1pt\hss\vrule height 5.1pt}\hrule}}

\begin{document}

\title{Quantum stochastic differential equation is unitary equivalent
to a symmetric boundary value problem in Fock space
\footnote {These paper is submitted to "Infinte Dimensional
Analyses  and Quantum Probability" and published partially in 
\cite{C96, C96M, C97}}}
\author{Alexander M.~Chebotarev}
\date{ }
\maketitle
\centerline{\tt 119899 Moscow, MSU, Quantum Statistics Dep.}
\centerline{\tt 109028 Moscow, MIEM, Applied Mathematics Dep.}
\bigskip

\begin{abstract}{\sl
       We show a new remarkable connection between the symmetric
       form of a quantum stochastic differential equation (QSDE)
       and the strong resolvent limit of Schr\"odinger equations
       in Fock space: the strong resolvent limit is unitary equivalent to
       QSDE in the adapted (or Ito) form, and the weak limit
       is unitary equivalent
       to the symmetric (or Stratonovich) form of QSDE.

      We prove  that QSDE
	is unitary equivalent to a symmetric bo\-un\-da\-ry value problem
      for the Schr\"odinger equation in Fock space. The boundary
      con\-di\-ti\-on describes standard jumps of
      the phase and amplitude of components
      of Fock vectors belonging to the range of the resolvent. The corresponding
      Markov evolution equation (the Lindblad or Markov master equation)
      is derived from the boundary
      value problem for the Schr\"odinger equation.}

\end{abstract}
\bigskip

\section{Introduction}
          The last two decades show a valuable progress in
          quantum probability
          theory and applications \cite{AFL}--\cite{Pa}. It was
          discovered that
          fundamental constructions of  classical probability
          theory, such as  central limit theorems,
          conditional expectations, martingales, stopping times, the
          Markov property, and Markov
          evolution equations, have noncommutative generalizations
          \cite{Pa}--\cite{Me}.
          The quantum stochastic differential equation (QSDE)
          is a noncommutative generalization of the Ito stochastic
          equation
          suitable for describing irreversible Markov evolution
          in operator algebras
          \cite{ALV}--\cite{ZG}. The QSDE and the
          Schr\"odinger equation describe a unitary
          evolution of a quantum system and its environment.
          The solution of the Schr\"odinger equation is  a one-parameter
          unitary group $U_t$, and the solution of the QSDE is
           a unitary cocycle  $u(s,t)$,
          i.e. an interval-dependent family of unitary operators
          with one of two composition laws
          $u(s,\tau)u(\tau,t)=u(s,t), \; s\le\tau\le t\;$
          for the  right cocycle, or
          $u(t,\tau)u(\tau,s)=u(t,s)$ for the left cocycle.
          From a mathematical viewpoint,
          this structure difference between $U_t$ and $u(s,t)$
          is superficial and physically unobservalble, because $u(s,t)$
          is  usually  an interaction representation of some
          unitary group  $U_{t-s}.$

       A deep distinction between a QSDE and a typical Schr\"odinger
       equation is that the QSDE necessarily involves a singular component
          in its formal Hamiltonian, i.e. it
   can be regarded, up to unitary equivalence, as a Schr\"odinger
          equation with  Hamiltonian operator
          perturbed by  singular
          bilinear forms (see \cite{K89}--\cite{AKK}).

          Such Hamiltonians in ${\cal  H}\otimes \Gamma^S(L_2(\r))$
          appear as strong resolvent
          limits ($r-\lim$ or $srs-\lim$) of self-adjoint operators
                  $\widehat H_{\alpha}=H\otimes I+I\otimes \widehat{\bf E}+
      H^{(\alpha)}_{int}$ depending on a scaling parameter $\alpha\in(0,1]$
      such that
$$
H^{(\alpha)}_{int}=R^*\otimes A(f_{\alpha})+
R\otimes A^+(f_{\alpha})+K\otimes A^+(g_{\alpha}) A(g_{\alpha}),
$$
   where $K=K^*>0,\; (R^*)^*=R,\;$ $f,g\in L_2(\R),\;$
   $f_{\alpha}(\omega)=f(\alpha\omega),\,
        g_{\alpha}(\omega)=g(\alpha\omega),\;$  $f(0)=g(0)=(2\pi)^{-1/2}.$
        The family of quadratic forms
        $H_{\alpha, *}[h\otimes\psi(v)]=(h\otimes\psi(v),\widehat H_{\alpha}\,h\otimes\psi(v))$
        converges clearly to the quadratic form
\begin{eqnarray*}
{H}_*[h\otimes\psi(v)]&=&
e^{||v||^2}\biggl( (h,Hh)+||h||^2 (\psi(v),\widehat{\bf E}\psi(v))\\
&+&(Rh,h)\widetilde v(0)+(h,Rh)\overline{\widetilde v(0)}+(h,Hh)|\widetilde v(0)|^2\biggr),
\end{eqnarray*}
       with the singular component
       vanishing on the total subset
       consisting of vectors $h\otimes\psi(v)
       \in{\cal  H}\otimes \Gamma^S(L_2(\r))$ such that
       $ v\in C_0^{\infty}(\r),\;  \widetilde v(0)=0,$
  $h\in{\rm dom\,}K\cap {\rm dom\,}R \cap{\rm dom\,}R^*\cap{\rm dom\,}H.$

We prove, for commuting coefficients $K,\;R,$ and $H$ that,
in the strong resolvent sense, $\widehat H_{\alpha}$ converges to
a quadratic form with another regular and singular components:
\begin{eqnarray*}
{\bf H}_*[h\otimes\psi(v)]&=&
e^{||v||^2}\biggl( (h,H_0h)+
||h||^2 (\psi(v),\widehat{\bf E}\psi(v))\\
&+&(h,H_1h)\widetilde v(0)
+ (h,H_2h)\overline{\widetilde v(0)}+(h,H_3h)|\widetilde v(0)|^2\biggr),
\end{eqnarray*}
where
$$
{ H}_0=H-R^* {K\over {4+K^2}}R+
R^*{2i\over{4+K^2}}R\stackrel{\rm def}{=}-iG,\quad { H}_1=
R^*{2\over{ 2-iK}}\stackrel{\rm def}{=}iL^*W,
$$
$$
{ H}_2\stackrel{\rm def}{=}{2\over{ 2-iK}}R=-iL,\quad
{ H}_3={2K\over{2-iK}}=i(I-W),\quad  W={2+iK\over2-iK}.
\eqno(1.1)
$$
         Ill-defined operators $A,\, A^+,$ and $\Lambda$ that
         correspond formally to singular quadratic forms
$$
A_*[\psi(v)]=e^{||v||^2}\widetilde v(0),\quad
A^+_*[\psi(v)]=e^{||v||^2}\overline{\widetilde v(0)},\quad
\Lambda_*[\psi(v)]=e^{||v||^2}|\widetilde v(0)|^2
$$
generate well-defined operator-valued measures
$$
M_1(T)=A(T)=\int_T\,dt\,J_t^*AJ_t,\quad
M_2(T)=A^+(T)=\int_T\,dt\,J_t^*A^+J_t,
$$
$$
M_3(T)=\Lambda(T)=\int_T\,dt\,J_t^*\Lambda J_t
$$
        which, up to unitary equivalence, act as fundamental
        {\it creation, annihilation and number}
        processes in the Hudson--Parthasaraty framework:
$$
A_*(T)[\psi(v)]=e^{||v||^2}\int_T\,dt\,\widetilde v(t),\quad
A^+_*(T)[\psi(v)]=e^{||v||^2}\int_T\,dt\,\overline{\widetilde v(t)},
$$
$$
\Lambda_*(T)[\psi(v)]=e^{||v||^2}\int_T\,dt\,|\widetilde v(t)|^2,
$$
        where $\widetilde v(t)={\cal  F}_{\omega\to t}v(\omega)$
        is the Fourier transform of $v(\omega).$
         Set $M_0(T)={\rm mes\,} T$ and denote by
         $M(T)=\sum_0^3\,L_k\otimes M_k(T)$  an
          operator-valued measure in the QSDE
$$
d\,u(0,t)=u(0,t)M(dt_+),\,\;M(dt_+)=M(t,t+dt).
$$
                   One of our main results explains a connection
          between the unitary group $U_t$ and the solution $u(s,t)$
          of  QSDE:
$$
U_t=e^{it\widehat {\bf H}}=s-\lim_{\alpha\to0}e^{it\widehat H_{\alpha}}
=u(0,t)J_t,
\eqno(1.2)
$$
$$
M(T)=\sum_0^3\,L_k\otimes M_k(T)=i\int_T\,d\tau (J_{\tau}\widehat{\bf H}J^*_{\tau}-
I\otimes\widehat{\bf  E})\qquad
\forall T\in {\cal  B}(\R),
$$
        where $L_k=iH_k$  (see equation (1.1)), and
         $J_t=I\otimes e^{it\widehat{\bf  E}},
        \;\widehat{\bf  E}=\int\,d\omega\,\omega\,a^+(\omega)\,a(\omega)$
        is an
        environment energy operator generating
        an interaction representation of the unitary evolution
$$
u(s,t)=J_sU_{t-s}J^*_t.
$$

        A remarkable observation is that the weak limit
        $\widehat H= w-\lim \widehat H_{\alpha}
        \neq\widehat {\bf H}=r-\lim \widehat H_{\alpha}$
        can be derived by a symmetrization of the QSDE:
        $u(0,t)M(dt_+)=u(0,t)N(dt),$
        where $M(dt_+)=M(t,t+dt)$ is the stochastic differential
        of the adapted equation,
         and $N(dt)=N(t-dt/2,t+dt/2)$ is the stochastic differential of
         the corresponding symmetric equation.
        A relation between the symmetric differential $N$
        and the corresponding adapted differential $M$, proved in \S2,
        has the form of an integral equation
$$
N(T)+{1\over2}\int_T\,M(dt_+)N(dt_+)=M(T)\quad
\quad \forall T\in{\cal  B}(\R),
\eqno(1.3)
$$
        and an unexpected fact is that the weak limit
        $\widehat H$ contributes to the measure  $N(T)$ of
         the symmetric differential:
$$
N(T)=i\int_T\,d\tau (J_{\tau}
\widehat H J^*_{\tau}-I\otimes \widehat{\bf  E})\quad
{\rm or }\quad  \widehat H=I\otimes \widehat{\bf  E}+
{1\over i{\rm\, mes\,}T} \int_T\,J^*_{\tau}N(d\tau) J_{\tau}.
$$
        Equation (1.3) implies the Hudson--Parthasaraty
        necessary condition for a solution of
        QSDE to be unitary  \cite{HP}:
$$
L_0=iH_s-{1\over2}L^*L,\quad
L_1=-L^*W,  \quad
L_2=L,\quad
L_3=W-I
$$
        where $L$ and $W$ are related to operators $K$ and $R$ as in (1.1),
        and $H_s=H-{i/2}L^*(I-W)(I+W)^{-1}L=H-R^*K(4+K^2)^{-1}R.$
        We would like to remark that the solution of
        (1.3) differs from solutions of the equation
        $M(dt)=\exp\{N(dt)\}-I$ derived in \cite{H} for
        the symmetric operator-valued measure $iN(T).$

       Very technical  assumptions sufficient for the unitary  property of
       solutions of QSDE
       were obtained in the last decade basically by perturbation
        methods \cite{ALV},  \cite{H}--\cite{BFS}, but symmetric operators
responsible for the unitary property of solutions of QSDE were not discovered.
        The difficulties are connected with the
         violation of the group property by solutions of QSDE (see \cite{J}),
         and with the violation of the symmetric property of the
         form-generators considered on the domain of Fock vectors with smooth components.
         More precisely (see \cite{C96}), the formal generator of
         the Schr\"odinger equation, which is unitary equivalent to QSDE,
         reads~as a dissipative operator, perturbed by
         nonsymmetric singular (in the sense of
         \cite{K89}--\cite{AKK}) bilinear form.

        In this paper  we consider a class of QSDE which appears
         as an interaction representation of the strong resolvent
         limit of the Schr\"odinger equations
         $\partial_t\psi_t=i\widehat H_{\alpha}\psi_t$
          in the Fock space parameterized by the scaling variable $\alpha.$
        The basic property of the limit
         resolvent is the existence of standard { \it
         jumps of amplitude and phase} of Fock vector components belonging to
         ${\cal  R}$:
$$
(\widehat N+1)^{-1}\bigl(I\otimes A(\delta_+)-W\otimes A(\delta_-)\bigr)\Psi=
(L\otimes I)\,\Psi,\quad\forall \Psi\in{\cal  R}
\eqno(1.4)
$$
        where $W=(2+iK)/(2-iK)$, $L=2i/( 2-iK)\,R,$
        $\Psi=\{\Psi_0,\dots,\Psi_n(\omega_1,\dots,\omega_n),\dots\},$
$$
A(\delta_{\pm})\widetilde\Psi_n(\tau)
=\lim _{\varepsilon\to\pm0}
\sum_{k=1}^n \widetilde{\Psi}_{n+1}
(\tau_1,\dots,\tau_{k-1},\epsilon,\tau_{k},\dots,\tau_{n}),\quad
\widehat N\Psi_n(\omega)=n\Psi_n(\omega),
$$
        $\widetilde\Psi_{n}(\tau_1,\dots,\tau_n)$ is the Fourier transform
         ${\cal  F}_{\omega\to\tau}$ of n-th component of
       Fock vector $\Psi.$

         The support of discontinuities of vectors from
         {\cal  R} coincides with  the support of
         singularities of the quadratic form  associated to the formal
         generator of the limit unitary group. Our main observation
          is that the
         generator of group derived on ${\cal  R}$
         (i.e. in the set of Fock vectors with the standard discontinuity
         (1.4))
$$
\widehat{\bf H}=H_0\otimes I+I\otimes
\widehat {\bf E}+iL^*W\otimes A(\delta_-),\quad
\widehat {\bf E}={\cal  F}_{\tau\to\omega}
        \int_{\R\setminus\{0\}}d\tau a^+(\tau)a(\tau)\,i\partial_{\tau}
        {\cal  F}_{\omega\to\tau}^{-1},
\eqno(1.5)
$$
          is a symmetric operator, that is
          $(\Phi,\widehat {\bf H}\Psi)=(\widehat {\bf H}\Phi,\Psi)\;
          \forall \Psi,\Phi\in {\cal  R};$ its
         symmetric property can be verified independently under  assumptions
        less restrictive  than used
         for an explicit construction of the resolvent.

        The algebraic computations permit to derive
        the Markov master equation  directly from the boundary
        problem for  the Schr\"odinger equation.

        The structure of the paper is the following. First,
        in \S2  an algebraic analog of the Markov property  is introduced
         for operator--valued processes in the
        Fock space. The processes with this property are called
        {\it interval adapted}. It is proved prove that the
        assumption (1.3) is sufficient to transform  an adapted
        QSDE to the symmetric
        form. In \S3 we construct an explicit solution
        of  the QSDE with commuting coefficients and prove
        (1.1)--(1.2) for this particular case. A class of explicitly solvable
        Schr\"odinger equations in the Fock space  is analysed in \S4.
        A direct computation shows that QSDE in the weak
        Hudson--Parthasaraty
        form coincides with the interaction representation of
        the limit Schr\"odinger equation.  Next, in \S4 we consider
        the properties of the resolvent of the limit Schr\"odinger
        equation, and prove that it ranges in a set of Fock vectors
        with components that satisfy a standard discontinuity
        conditions (1.4). In \S5 the Markov master equation
        is derived from the boundary value problem for the limit
        Schr\"odinger equation. The main results of this paper
        were included in \cite{C96} and \cite{C97}.

        This work, in its final stage, was inspired by intensive
        discussions with Prof. L.~Accardi during a short stay of
        the author in  V.~Volterra
        Center in February 1996;  our interest in the problem
        of the parameterization of generators of QSDE by self-adjoint
        operators was initiated long ago by the papers
         \cite{AFL}, \cite{GC}, \cite{A}-\cite{Ac}.

\begin{sloppypar}
\section{Interval--adapted processes in  Fock space }
\end{sloppypar}
         Let ${\cal  H}$ and ${\cal  H}_E$ be Hilbert spaces.
         The Hilbert space of a quantum system and its environment
         has a form of the tensor product
           ${\H}={\cal  H}\otimes\Gamma^S,$ where ${\cal  H}$
           describes states of the quantum system, and
           $\Gamma^S= \Gamma^S({\cal  H}_E)$ is
           the symmetric Fock space describing a state of the environment.
           We denote by $(\cdot,\cdot),$ $||\cdot||$ a scalar product and a norm
         in the corresponding Hilbert space. The scalar product in $\Gamma^S$
           is generated by the scalar product in ${\cal  H}_E:$
$$
(F,F')=\overline F_0\,F'+\sum {1\over n!}(F_n,F'_n),\quad
F=\{F_0,F_1,\dots \}\in \Gamma^S ,\quad F_0\in \C,\quad
F_n\in \otimes_1^n {\cal  H}_E.
$$
           Let ${\cal  E}\subset \Gamma^{S}$  be a total subset
           of coherent (exponential) vectors   $\psi(f)$
        $=\{1,\,f,\,f\otimes f,\,\dots ,f\otimes \dots \otimes f,\,\dots\},$
           $f\in {\cal  H}_E$  (see \cite{Me}, \cite{Pa}).
            A symmetric tensor product is defined
            on the total subset ${\cal  P}$ of polinomial  vectors
           $F_j=f^{\otimes j}$ and  $F_{n-j}'={f'}^{\otimes n-j}$
           by the equation
$$
F \otimes F'=\biggl\{F_0\cdot F_0',\;\; F_0\otimes  F_1'+F_0'\otimes  F_1,
\;\;\dots,\sum_{j=0}^n\,
{\bf S}_n \bigl(F_j\otimes F_{n-j}'\bigr),\;\dots \biggr\},
\eqno(2.1)
$$
        where ${\bf S}_n$ is a sum over all permutations
        of arguments in multipliers $f,\,f'.$
                This definition can be extended by continuity to
        the linear span of  ${\cal  P}$ and to
        $\overline {{\rm Span\,} {\cal  P}}=\Gamma^S.$

       Definition (2.1) implies the relation
       $\psi(f)\otimes \psi(v)=$ $\psi(f+v)$ for coherent vectors.
       Hence, $\psi(v)=\otimes_i \psi(\widehat\pi_{T_i}v)$
       for any (finite) identity decomposition or
       complete system of projectors
       $\{\widehat\pi_{T_j}\}_{T_j\in{\cal T}}\subset {\cal  B}({\cal  H}_E),$
       parameterized by subsets of a measurable space
       $({\cal  T},\T):$
$$
\widehat\pi_T^*=\widehat\pi_T=\widehat\pi_T^2,\quad \forall T\in {\cal  T},
\quad  \widehat\pi_{T_1}\widehat\pi_{T_2}=
\widehat\pi_{T_2}\widehat\pi_{T_1}=0,\quad
\forall T_1,T_2:\, T_1\cap T_2=\emptyset.
$$
        The Fock space $\Gamma^S$ has a property of the
        tensor decomposition for any finite complete system of projectors.
        In particular,
$
\Gamma^S({\cal  L}_2(\R))=
\otimes _j\Gamma^S_{T_j},\quad \Gamma^S_{T_j}=
\Gamma^S({\cal F}{\cal  L}_2(T_j))
$
        for any finite disjoint partition $\{T_j\},$  and
        $\pi_T={\cal F}^*\,I_T\,{\cal F}$  for any unitary operator
       ${\cal F},$ where  ${\cal  T}=\R,\;{\cal  H}_E={\cal  L}_2(\R); \,$
        $I_T$ stands for an multiplication  operator  by the indicator
        function of a measurable subset $T.$

         Later $\H_{\cal T}^{\pi}$  denotes a product
        ${\cal  H}\otimes\Gamma^S$ equipped by the system of orthogonal
        projectors $\{\widehat\pi_T\}_{T\in{\cal  T}},$ and
        $\H_T$ is the subspace ${\cal  H}\otimes\Gamma^S_{T},$
        where $\Gamma^S_{T}=$ $\Gamma^S(\pi_{T}{\cal  H}_E).$
        The operator-valued family
        $u(\cdot):$ ${\cal T}\to{\cal  B}(\H)$ is called
        $(\widehat\pi,{\cal  T})$-{\it adapted}
        if for any $T\in {\cal  T}$ the following two assumptions hold:
$$
u(T)h\otimes \psi(\widehat\pi_{T}v)\in {\cal  H}\otimes\Gamma^S_{T},
\quad u(T) h\otimes \psi(v)=\psi(\widehat\pi_{T^a}v)\otimes\{ u(T) h\otimes
\psi(\widehat\pi_{T}v)\},
$$
        where $T^a=\R\setminus T;$ that is the operators
        $u(T)$ leave invariant  vectors from $\H_{T^a}
        \in \H_{\cal T}^{\pi}$  (and
        the subset $\H_{T}=$ ${\cal  H}\otimes\Gamma^S_T$).
         In what follows the Fourier transform is used as
         a unitary operator ${\cal F},$
         and the corresponding representation of multiplication
         by indicator functions play the role of
         projectors. In that case we call
         the family of operators  $\{u(T)\},\;T\in {\cal  B}(\R)$
         {\it interval adapted} (see \cite {C92}):
$$
(\varphi\otimes \psi(f),\,u(T)\,g\otimes \psi(v))=
\exp\{(f,\widehat\pi_{T^a}v)\}
(\varphi\otimes \psi(\widehat\pi_T\,f),u(T)\,
g \otimes \psi(\widehat\pi_T\,v)),
\eqno(2.2)
$$
	with $\varphi,g\in {\cal H},\quad \psi(f),\psi(v)\in {\cal E},$
        $ T^a=\R\setminus T.$  Definition (2.2) means that the operator $u(T)$
        changes  only the part of exponential vector from  $\Gamma^S_T.$
        Denote by ${\cal  B}({\H}_T)
        ={\cal  B}({\cal  H})\otimes {\cal  B}(\Gamma^S_T)$
        an algebra of all bounded operators acting as in (2.2).
        Equation (2.2) can be verified in the weak topology, and hence
        it holds for the adjoint operator $u^*(T)$ as a hereditary property of $u(T).$
        Relation (2.2) implies the commutativity of
        the components from $\Gamma^S_{T^a}$ of a coherent vector
        and interval adapted operator families $u(T),\,M(T):$
$$
(\varphi\otimes  \psi(f),u(T_1)\,M(T_2)\,g\otimes  \psi(v))=
\exp {(f,\widehat\pi_{T^a}v)}
$$
$$
\times\Bigl(\bigl\{u^*(T_1)\,\varphi\otimes
\psi(\widehat\pi_{T_1}\,f)\bigr\}\otimes
\psi(\widehat\pi_{T_2}\,f),\bigl\{M(T_2)g \otimes
\psi(\widehat\pi_{T_2}\,v)\bigr\}
\otimes \psi(\widehat\pi_{T_1}\,v)\Bigr),
\eqno(2.3)
$$
        where $\quad T_a=\R\setminus\{\cup T_i\},$ $\,T_1\cap T_2
        =\emptyset.$ An important example of interval adapted
        operators is {\it creation, annihilation} and
        {\it conservation} processes in the Fock space:
$$
(\psi(f),A(T)\psi(v))=e^{(f,v)}(\widetilde I_T,v),\quad
(\psi(f),A^+(T)\psi(v))=e^{(f,v)}(f,\widetilde I_T).
$$
$$
(\psi(f),\Lambda(T)\psi(v))=(f,\widehat\pi_Tv)\exp\{(f,v)\},
\eqno(2.4)
$$
         with $\widehat\pi_T={\cal  F}^*_{t \to \omega}
         I_T(t){\cal  F}_{\omega \to t}.$

        Let  $a^+(\omega),\,a(\omega),\;\omega \in \R$ be
        creation and annihilation operator densities,
        $\widehat {\bf E}=\int\,\omega \,a^+(\omega)\,a(\omega)\,d\omega$
        $\in {\cal  C}(\Gamma^S)$ (${\cal  C}(H)$ denotes a set of {\it
        closed} operators in a Hilbert space $H$)
        be a generator of the Journ\'e unitary transform:
        $J_t=\exp\{it\widehat {\bf E}\}.$
        The group of unitary operators $J_t\psi(v)=
        \psi(e^{i\omega t} v)$ is uniquely defined, and from the canonical
        commutation relations we have:
$$
a_t^+(\omega)=J_t^*\,a^+(\omega)\,J_t= e^{-i\omega t}a^+(\omega),\quad
J_t\,a^+(\omega)\,J_t^*= e^{i\omega t}a^+(\omega).
\eqno(2.5)
$$
       Let $\widetilde W^p_2(\R)$ be the Sobolev space with the scalar product
       $\langle f,v\rangle = (f(\omega), v(\omega) (1+\omega^2)^{p/2}),$
       $p\in\R.$ Denote by $A=A(1)$ $=\int_{\R}\, a(\omega)\,d\omega,$
        $A^+=A^+(1)$ $=\int_{\R} a^+(\omega)\,d\omega,$ and $\Lambda=A^+ A,$
        quadratic forms in $\widetilde W_1^2(\R)$, so that
        ${(\psi(f),A^+A\psi(v))}={2\pi}\,
        \overline {\widetilde f(0)}\widetilde v(0)e^{(f,v)},$
$$
{(\psi(f),A\psi(v))}=\sqrt{2\pi}\,\widetilde v(0)e^{(f,v)},\qquad
{(\psi(f),A^+\psi(v))}
=\sqrt{2\pi}\,\overline {\widetilde f(0)}e^{(f,v)}.
$$
        The quadratic forms $A,A^+,\Lambda$
        are not closable in $\Gamma^S({\cal L}_2(\R))$   and
        vanish on the dense set generated by coherent vectors
         $\{\psi(v):\widetilde v(0)=0\}.$
        They are closable in
        $\Gamma^S(\widetilde W^1_2(\R))$ (\cite{K89, K93})
        and define closed operators from $\Gamma^S(\widetilde W^1_2(\R))$
        to $\Gamma^S(\widetilde W^{-1}_2(\R)).$ Operators (2.4) can be
        expressed as operators, corresponding to densely defined closable
        quadratic forms:
$$
A(T)={1\over\sqrt{2\pi}}\int_{T}\,A_t\,dt,\quad
A^+(T)={1\over\sqrt{2\pi}}\int_{T}\,A^+_t\,dt,\quad
\Lambda(T)={1\over2\pi}\int_{T}\,A^+_tA_t\,dt
\eqno(2.6)
$$
        with $A_t=J_t^*\,A\,J_t, \;A_t^+=J_t^*\,A^+\,J_t$ and
        $\Lambda_t=J_t^*\,\Lambda\,J_t.$
         In the standard notation, the operator $A(T)$ coincides with the
         annihilation operator $A(\widetilde {I}_T),$ where
         $\widetilde {I}_T(\omega)$
        is the Fourier transform of the indicator function
        of a bounded set  $T\in {\cal  B}(\R).$

        Let $M(T)=\sum\,L_j\otimes M_j(T)$ be an additive function
        so that
$$
M_0(T)=I\cdot {\rm mes\,} T,\quad
M_1(T)=A(T),\quad  M_2(T)=A^+(T),\quad M_3(T)=\Lambda(T),
\eqno(2.7)
$$
        $L_j\in {\cal  C}({\cal  H}),\, M_j(T)\in {\cal  C}(\Gamma^S).$
        Assume there exist a joint dense domain $D_0\in{\cal  H}$ for $L_j$
        and their products.    Let $g\in D_0,\;h\in {\cal  H},\,|\widetilde f|,\,
        |\widetilde v|\le 1.$
        Identities  (2.4) generate the equation for correlators
$$
(h\otimes \psi(f),M(T)\,g\otimes \psi(v))= e^{(f,v)}\cdot\sum_{i=0}^3
(h,L_i g)\cdot((\widetilde f)^{\alpha_i},I_T\,\widetilde v^{\beta_i})=
O({\rm mes\,} T)
\eqno(2.8)
$$
        with $\alpha_i=1$ for  $i=2,3$  and $\beta_i=1$  for
         $i=1,3;$  $\alpha_i=\beta_i=0$ in other cases.
        It follows from here  that the family $M(T)$ is interval
        adapted in the sense of definition (2.2), and that the mean
        value  is absolutely continuous with respect to the standard
        Lebesgue measure:
$$
{d\over dt}\bigl(h\otimes \psi(f),M(0,t)\,g\otimes \psi(v)\bigr)
= e^{(f,v)}\cdot\sum_{i=0}^3(h,L_i g)
\overline {(\widetilde f(t))^{\alpha_i}}\widetilde v (t)^{\beta_i}\in L_1(\R).
\eqno(2.9)
$$
        These observations are important for a rigorous  definition
        of the QSDE in the weak form \cite{HP}.

         Consider the operator-valued stochastic differential equation
$$
du(\tau,t)=u(\tau,t)\,M(dt_+),\quad  \tau\le t,
\quad s-\lim_{T\to \emptyset}\, u(T)=I
\eqno(2.10)
$$
        (see \cite{HP}), where
           $M(T)$ is an additive function on ${\cal  B}(\R)$ which ranges
           in the set of interval adapted closable operators
           wth a common dense domain ${\cal  D}\subseteq \H.$
         The strongly continuous  interval adapted cocycle
           $u(\cdot):{\cal  B}(\R)\to {\cal  B}(\H)$
           is called {\it a weak solution} of equation (2.10) if for any
           $\varphi\in {\cal  D}$ there exist a limit
$$
\int_T\,u(s,\tau)M(d\tau_+)\varphi=\omega-\lim_{N\to\infty}
\sum_{j\in J_N(T)}u(s,t_j)M(t_j,t_{j+1})\varphi,
$$
            so that
$
u(s,t)\varphi=\varphi+\int_s^t\,u(s,\tau)M(d\tau_+)\varphi,
$
        with  $t_j=j2^{-N},\; J_N(T)=\{j:\;t_j\in T \}.$
          From the bound (2.9) follows an existence of time derivatives
            of the cocycle $u$
$$
{d\over dt}\bigl(h\otimes\psi(f),u(\tau,t)g\otimes\psi(v) \bigr)=
\sum_i\biggl(h\otimes\psi(f),u(\tau,t)L_ig\otimes\psi(v) \biggr)
\overline {\widetilde f(t)^{\alpha_i}}\widetilde v (t)^{\beta_i}
\eqno(2.11)
$$
           $h\in{\cal  H},\;g\in D_0,\;f,v\in {\widetilde W}_2^1(\R).$
           Equation (2.11) is called {\it a weak}
           form of QSDE (2.10). In this section
           we aim to derive a symmetric form of the equation
         (2.10), that is to construct an operator-valued measure
        $N(T)$ such that
$$
w-\lim\,\varepsilon^{-1}\,u(t)\,M(t,t+\varepsilon)\varphi
=w-\lim\,\varepsilon^{-1}\,
u(t)\,N(t-{\varepsilon/2},t+{\varepsilon/2})\varphi \quad \forall \varphi\in
{\cal  D}.
$$
         We start with a suitable choice of operator-valued
         measures $M(\cdot)$ in equation (2.10).
                  Let ${\cal  D}$ be a total  subset in
           ${\H}=\H^{\pi}_{\cal  T},$ ${\cal  T}={\cal  B}(\R),$
           $\widehat\pi_T={\cal  F}^*_{\tau\to\omega}I_T(\tau)
           {\cal  F}_{\omega\to\tau}.$ The main property of the projector
           $\widehat\pi_T$ follows from the properties of the
           Fourier transform:
$$
e^{i\omega t}\widehat\pi_T e^{-i\omega t} =\widehat\pi_{T+t},\quad
{\cal  F}_{\omega\to t}v=(2\pi)^{-1/2}\int_{\R}\,d\omega\,
e^{-i\omega t}v(\omega),\quad
T\in {\cal  B}(\R),\quad t,\omega\in \R.
$$
\vskip2mm
\par\noindent
           {\bf Example 2.1.} For equation (2.10) with bounded
           coefficients $\{L_j\}$ we put
           ${\cal  D}={\cal  H}\otimes {\cal  E}_1,$ where
           ${\cal  E}_1\subset \Gamma^S$ is a subset of
        coherent vectors $\psi(f)$ such that
        $|\widetilde f|\le 1,$ and
        $f\in {\cal  L}_2(\R)$ is a finitely supported function.
        ${\cal  E}_1$ is total in $\Gamma^S$ (\cite{Me}) and
        ${\cal  D}$ is total in $\H.$ It is convenient to put ${\cal  D}={D}_0\otimes {\cal  E}_1,$
        for equations with  unbounded coefficients, if there exist
        where  is a dense joint domain ${D}_0\subseteq{\cal  H}$
        for all $L_k$ and their products.
\vskip2mm
\begin{definition}
                 The vector subspace of
                additive interval adapted functions
                  $M:{\cal B }(\R)\to {\cal  B}(\H^{\pi}_{{\cal  T}})$
                  is labelled by ${\cal  A}_{{\cal  D}}$ if
                  for any  $\varphi\in{\cal  D}$
\begin{itemize}
\item[(\rm{i})]           there exist an upper bound
$$
\sup\limits_{\{T_j\},\{\psi_j\}}^{}
\sum_j\,|(\psi_j,M(T_j)\varphi)|={\mu}_{T}(\varphi),
\eqno(2.12)
$$
         where $T$ is a bounded Borelean subset of $\,\R,$
          $\{T_j\}$ is a disjoint partition of $T$,
          $\{\psi_j\in \H_{T_j^a}\}$ is a uniformly
          bounded family of vectors from $\H,$
          $\sup_j||\psi_j||\leq 1,$
        and $\mu_T$ is a finite measure such that
        $\mu_T={\mu}_{T}(\varphi)=0$ if ${\rm mes}\,T=0;$
\item[(\rm{ii})] there exist a derivative
$$
f_t(\psi,\varphi)={(\psi,M(dt_+)\varphi)\over dt}=
\lim_{\varepsilon \to 0}\varepsilon^{-1}\,(\psi,M(t,t+\varepsilon)\,\varphi),
\eqno(2.13)
$$
        which is a continuous function from
        $\H \to {\cal  L}_1^{loc}(\R)$ for any fixed $\varphi\in{\cal  D}.$
\end{itemize}
\end{definition}
           Assumption (i) requires an existence of the
           weak absolute upper bound for  integral sums
           for every natural $N$ and partitions $\{T_j\};$ condition
           (ii) assumes an existence of the majorizing function
           $(\psi,M(T)\varphi)$
           which is absolutely continuous w.r.t. the standard Lebesgue
           measure.

        To fix a notation  set   ${\widehat\pi_T={\cal  F}_{t\to\omega}
        I_T{\cal  F}^*_{\omega\to t}}$ as a system of projectors,
         and ${\cal  B}(\R)$ as a Borelean $\sigma$-algebra ${\cal  T}.$
        Denote $\delta_N=2^{-N}=\tau_1,\,
        \tau_j=\tau_j^{(N)}=j\,2^{-N},\quad J_N(T)=\{j:\,\tau_j^{(N)}\in T\}.$
\begin{lemma}   For every $L\in {\cal  A}_{{\cal  D}},\;\varphi\in{\cal  D}$ and
        the strongly continuous interval adapted bounded cocycle
        $u(t)=u(0,t)$ there exist an interval adapted weak limits
$$
\lim_{N\to \infty} \sum_{j\in J_N(T)}
\,u(\tau_j^{(N)})\,
L(\tau_j^{(N)},\tau_{j+1}^{(N)})\varphi
=\int_T\,u(\tau)\,L(d\tau_+)\varphi,
\eqno(2.14)
$$
$$
\lim_{N\to \infty}\,
2\sum_{j\in J_N(T)}u(\tau_j^{(N)})
\,L(\tau_j^{(N)},\tau_j^{(N)}+\delta_{N+1})\varphi=
\int_T\,u(\tau)\,L(d\tau_+)\varphi,
\eqno(2.15)
$$
        called  adapted  stochastic integrals.
\end{lemma}
 \vskip3mm
 \par\noindent {\bf Proof}.
         We can apply the Lebesgue theorem on the dominated convergence
        to prove the equation (2.14) because there exist a uniform
        absolute bound (2.12)  for sums:
$$
J_N=\sum_{j\in J_N(T)} |(\psi,u(\tau_j)\,L(\tau_j,\tau_{j+1})\,\varphi)|
\le\sup_N \sum_{j\in J_N(T)}|(\psi_j,L(\tau_j,\tau_{j+1})\,\varphi)|
\le {\mu}_{T}(\varphi),
$$
	where $\psi_j=u^*(0,\tau_j)\,\psi,$ and the limit
$
f_{\tau}(\psi,u^*(\tau)\,\varphi)=
\lim\,\varepsilon^{-1}\,
(u^*(\tau)\,\psi,L(\tau,\tau+\varepsilon)\,\varphi)
$
        belongs to $L_1^{loc}$ because of (2.13).
        Hence, $\lim J_N = J_T(\psi,\varphi)=
        \int_T\,d\tau\, f_{\tau}(u^*(\tau)\,\psi, \varphi).$
        Since $f_{\tau}(u^*(\tau)\,\cdot,\varphi):\H\to L_1^{loc}(\R)$ is
        a continuous function for any $\varphi\in{\cal  D},$ there exist
        a unique element $\varphi_T\in\H$ so that $(\psi,\varphi_T)=
        J_T(\psi,\varphi).$
        Clear that $\varphi_T:{\cal  B}(\R)\to\H$ is an additive function,
        and $(\psi,\varphi_{T_n})\to 0$ for any decreasing sequence $\{T_n\}$
        such that $\cap_{n\ge 1} T_n=\emptyset$ because of
        (2.13). Therefore, $\varphi_T$ is a vector-valued measure, and we
        write $\varphi_T=\int_T\,u(\tau)\,L(d\tau_+)\varphi.$  Thus, we prove
        (2.14).

        To prove (2.15) consider an identity
$$
2\sum_{j\in J_N(T)}u(\tau_j^{(N)})
\,L(\tau_j^{(N)},\tau_j^{(N)}+\delta_{N+1})\varphi=
\sum_{j\in J_N(T)}u(\tau_j^{(N)})
\,L(\tau_j^{(N)},\tau_{j+1}^{(N)}+\delta_{N+1})\varphi+
$$
$$
+\sum_{j\in J_N(T)}u(\tau_j^{(N)})
\,\biggl(L(\tau_j^{(N)},\tau_{j}^{(N)}+\delta_{N+1}) -
L(\tau_{j}^{(N)}+\delta_{N+1},\tau_{j+1}^{(N)})\biggr)\varphi.
$$
        As we noticed earlier, the first sum converges to integral (2.14).
        The second sum is absolutely  bounded, and each term
        converges  weakly to  0. Hence, the Lebesgue theorem on dominated convergence
        implies a trivial limit for
         the second summand as $N\to \infty.$ \qed

        Let $M^{(1)},M^{(2)},\dots\in{\cal  A}_{{\cal  D}},$  for simplicity
        one can think that
        $M^{(k)}(T)=L_k\otimes M_k(T)$ (see (2.7)).
\begin{definition} Locally convex vector space
$G_{\cal  D}=G_{\cal  D}(M^{(1)},M^{(2)},\dots)\subseteq {\H}$ denotes a
          closure of
   ${\rm Span\,}{\cal  D}\subseteq\H,$ equipped by seminorms:
$$
\rho_t(\varphi)=||\varphi||+
\sup_{\tau,\,\psi,\,\{i_k\},\,\{T_j\}}{|(\psi,M^{(i_1)}(T_1)M^{(i_2)}(T_2)
\,J_{\tau}\,\varphi)|\over \prod_j \,{\rm mes}\, T_j}
\eqno(2.16)
$$
        with $T_1\cap T_2=
        \emptyset,\;\psi\in\H_{T^a},\;||\psi||\leq 1,\,$
        $T_a=\R\setminus \{T_1\cup T_2\},$
        $T_j\in(0, t),\;t\in\R_+.$ We assume that $G_{\cal  D}$ is dense in $\H.$
\end{definition}

        The group $J_{\tau}$ leaves invariant the set $G_{\cal  D}:$
        $\rho_t(g)=\rho_t(J_{\tau}g),$ $G_{\cal  D}=J_{\tau}G_{\cal  D}.$  Cocycle property
         of coefficients (in the sense of paper \cite{J})
         is a simple consequence of the definition  of $J_t:$
$
J_t^*A(T)J_t\psi(v)=(\widetilde I_T,e^{i\omega t}v)\psi(v)=
        (\widetilde I_{T+t},v)\psi(v)=A(T+t)\psi(v),\,
        J_t^*A^+(T)J_t\psi(v)=A^+(T+t)\psi(v).
$
        This property of the coefficients $M(\cdot)$ of equation (2.10) generates
        a cocycle property of the solution $u(T).$
         Later we will assume that the measures
         $M(\cdot)\in {\cal  A}_{{\cal  D}}$ poses the same property:
$
M(T+t)=J^*_{t}\, M(T) \,J_{t}.
$
        It is clear that $G_{\cal  D}$ is dense
        in $\H$  under assumptions of Example 2.1.
\begin{theorem} Let the interval adapted cocycle
                $u(t)$ be a bounded weak solution of (2.12),
                $\varphi\in {\cal  D}$ and
                $L ,\,ML,\,M^2\in {\cal  A}_{{\cal  D}}.$
         If the semi-norm $\rho_{\tau}(L(0,\tau)\,\varphi)$ (2.16) is
         uniformly bounded for all $\tau\in\R_+,$ then the
         adapted and the symmetric stochastic integrals satisfy the
         equation
$$
\int_T\, u(\tau)\,\biggl(L(d\tau_+)+
{1\over2}\,M(d\tau_+)L(d\tau_+)\biggr)\varphi
=\int_T\, u(\tau)\,L(d\tau)\varphi\quad
\eqno(2.17)
$$
        for every bounded set $T\in{\cal  B}(\R).$
\end{theorem}
         {\bf Proof.} Let $u(t)$ be a weak interval adapted solution
         of the equation (2.12). Consider a family of operators
         $u_s(t)=u(s)+u(s)M(s,t).$
         The difference $\Delta_u(s,t)=u(t)-u_s(t)$
         satisfies the equation
$
\Delta_u(s,t)=
\int_s^t\int_s^{\sigma}\,u(\tau)\,M(d\tau_+)\,M(d\sigma_+)
$
        in the weak sense, and for every $\psi\in\H,
        \;\varphi\in  {\cal  D}$ there exist a uniform bound
$$
|(\psi,\Delta_u(s,t)\,L(\tau_j^{(N)},\tau_{j+1}^{(N)})\,\varphi)|\le
{(t-s)^2\over2}\sup_{\tau\in T}
||u^*(\tau)\psi||\,\rho_{t-s}(L(0,2^{-N})\,\varphi).
\eqno(2.18)
$$
        Since $L(\cdot)$ is an additive function, from the equation
        for $u(t)$ we have
$$
u(\tau_j)\,L(\tau_j,\tau_{j+1})\varphi=\biggl\{u(\tau_{j+{1\over2}})L(\tau_j,\tau_{j+1})
-u(\tau_j)M(\tau_j,\tau_{j+{1\over2}})L(\tau_j,\tau_{j+{1\over2}})-
$$
$$
-u(\tau_j)\,M(\tau_j,\tau_{j+{1\over2}})\,L(\tau_{j+{1\over2}},\tau_{j+1})-
\Delta_u(\tau_j,\tau_{j+{1\over2}})L(\tau_j,\tau_{j+1})\biggr\}\varphi.
\eqno(2.19)
$$
        For every $\psi\in\H,\;\varphi\in {\cal  D}$
        a prior bound  follows  from the definition (2.16)
$$
\delta^{(1)}_{j,N}=
|(\psi,u(\tau_j)\,M(\tau_j,\tau_{j+{1\over2}})\,
L(\tau_{j+{1\over2}},\tau_{j+1})\,\varphi)|\le
\tau_1^2\sup_{\tau\in(0,t)}||u^*(\tau)\psi||\,\rho_{\tau_1}(\varphi),
$$
        and from (2.12), (2.16) we obtain
$$
\delta^{(2)}_{j,N}=
|(\psi,\Delta_u(\tau_j,\tau_{j+{1\over2}})M(\tau_j,\tau_{j+1})
L(\tau_j,\tau_{j+1})\,g)|
$$
$$
\le\tau_1^2\sup_{\tau\in(0,t)}||u^*(\tau)\psi||\,\rho_{\tau_1}(L(0,\tau_1))\,\varphi),
\eqno(2.20)
$$
        where  the semi-norm
        $\rho_{\tau_1}(L(0,\tau_1)\,\varphi)$ is uniformly bounded in $N$.
        Therefore,
        $$\lim_{N} \sum_{j}\,\delta^{(1,2)}_{j,N}=0.$$
        Now it is sufficient to observe that for operators
                $L,ML\in{\cal  A}_{{\cal  D}}$
        the weak convergence  in $\H$ take place   by Lemma 2.1:
$$
\lim_{N\to \infty} \sum_{J_N(T)}\,
u(0,\tau_j)\,L(\tau_j,\tau_{j+1})\varphi=
\int_T\,u(0,\tau)\,L(d\tau_+)\varphi,
$$
$$
\lim_{N\to \infty} \sum_{J_N(T)}\,
u(0,\tau_j)\,M(\tau_j,\tau_{j+{1\over2}})
L(\tau_j,\tau_{j+{1\over2}})\varphi=
{1\over2}\int_T\,u(0,\tau)\,M(d\tau_+)L(d\tau_+)\varphi.
$$
        Hence, the weak limit
        is well-defined on ${\cal  D}$ by identity (2.19)
        and is called {\it a symmetric} (in the Stratonovich sense)
        stochastic integral:
$$
\int_T\, u(\tau)\,L(d\tau)=
\lim_{N\to \infty}
\sum_{j\in J_N(T)}\,u(\tau_{j+{1\over2}})\,L(\tau_j,\tau_{j+1})
$$
$$
=\int_T\,u(\tau)\biggl(L(d\tau_+)+{1\over2}M(d\tau_+)L(d\tau_+)\biggr).
\eqno(2.21)
$$
        \qed

        Equation (2.21) is satisfied whenever the measure  $L(\cdot)$
        of the symmetric integral satisfies the   equation
$$
L(T)+{1\over2}\int_T \,M(dt_+)L(dt_+)=M(T) \qquad \forall T\in{\cal  B}(\R)
\eqno(2.22)
$$
        with  the given operator-valued measure $M(T)$
        in the adapted integral.
        Assume that
$$
M(T)=\sum_0^3{\cal L}_i\otimes M_i(T),
\quad L(T)=\sum_0^3\ell_i\otimes M_i(T),
$$
$$
\quad M_0(T)=I\cdot{\rm mes\,}T,\quad M_1(T)=A(T),\quad
M_2(T)=A^+(T),\quad M_3(T)=\Lambda(T).
$$
        In this case the equation (2.22) has the form
        $\int_T\,\bigl\{L(dt) + {1\over2} \,M(dt)L(dt)\bigr\} = M(T).$
        Using the Ito multiplication table for stochastic
        differentials (see \cite{HP})
\vskip3mm
\centerline{
\begin{tabular}{|l||l|l|l|l|}
\hline
{ }      & $dA^+$ & $dA$ & $d\Lambda$ & $dt$ \\
\hline
\hline
$dA^+$     & 0    &  0 &   0       & 0 \\
\hline
$dA$       & $dt$    &  0 & $dA$      & 0 \\
\hline
$d\Lambda$ & $dA^+$  &  0 & $d\Lambda$ & 0 \\
\hline
$dt$       &  0     &  0 &  0      & 0 \\
\hline
\end{tabular}
}
\vskip3mm
\noindent
         in the left part of the equation
$
L(dt)+{1\over2}M(dt)L(dt)=\sum{\cal L}_i\otimes M_i(T)
$
        we obtain a system of linear equations with respect to $\ell_k:$
$$
{\cal L}_0={\ell}_0+{\cal L}_1\ell_2/2,\quad
{\cal L}_1={\ell}_1+{\cal L}_1\ell_3/2,\quad
{\cal L}_2={\ell}_2+{\cal L}_3\ell_2/2,\quad
{\cal L}_3={\ell}_3+{\cal L}_3\ell_3/2.
$$
        Assume that the operator $2+{\cal L}_3$ is invertible, and
        $(2+{\cal L}_3)^{-1}:{\cal  H}\to {\rm dom\,}{\cal  L}_1.$
        Then the solution  of this system looks as follows:
$
{\ell}_3=2{\cal L}_3(2+{\cal L}_3)^{-1},\quad
{\ell}_1=2{\cal L}_1(2+{\cal L}_3)^{-1},\quad
{\ell}_2=2(2+{\cal L}_3)^{-1}{\cal L}_2,\quad
{\ell}_0={\cal L}_0-{\cal L}_1(2+{\cal L}_3)^{-1}{\cal L}_2.
$
        If we take
$$
{\cal  L}_0=iH+R^*{1\over2-iK}R=-G,\quad
{\cal  L}_1=-L^*W,\quad  {\cal  L}_2=L, \quad  {\cal  L}_3=W-I,
\eqno(2.23)
$$
        as the coefficients $\{{\cal  L}_k\}$
        satisfying the condition necessary for existence
        of a unitary solution of (2.1) (see \cite{HP}), then
$
{\ell}_3=iK,\;
{\ell}_1=iR^*,\;
{\ell}_2=iR,\;
{\ell}_0=iH,
$
        where $K=2i{I-W\over I+W},\;
        R=-{2i\over I+W}L$ are the coefficients
        of the symmetric operator $\widehat {H}(T):$
$$
i\widehat {H}(T)=\sum_{k=0}^3\ell_k\otimes M_k(T),
\quad M_k(T)=\int_T\,dt\,J_t^* M_kJ_t
\eqno(2.24)
$$
        with $M_0=I,\,M_1=A,\,M_2=A^+,\,M_3=A^+A.$

        Since $K$  is a self--adjoint operator, there exist a
        unitary operator $W=(1+iK/2)(1-iK/2)^{-1}$ called {\it a
        Cayley transform} of $K/2,$ and $2+{\cal L}_3=I+W=2(I-iK/2)^{-1}$
        is an invertible operator. Furthermore, the operator
        ${\cal  L}_1=iR^*(I-iK/2)^{-1}$ is well-defined
        provided the operators $K,\,R$ and $R^*$ have a joint dense domain $D_0;$
        the operators  ${\cal  L}_0=iH+R^*(2-iK)^{-1}R),$
         ${\cal  L}_2=(I-iK/2)^{-1}iR=-W{\cal  L}_1^*$ are well-defined also.
        We call $H,\,K,\,R,\,R^*$ {\it coefficients} of the interval-adapted
        measure $M(T)=\int_T\,dt\,\sum {\cal  L}_k\otimes J_t^*M_kJ_t,$
        and the operators $\{-i\ell_k\}$ are called  coefficients of the
        corresponding symmetric measure
        $\widehat {H}(T)=-i\int_T\,dt\,\sum {\ell}_k\otimes J_t^*M_kJ_t.$
        The assumptions on coefficients of the symmetric measure
        $\widehat {H}(T)$
        have a natural meaning: they are necessary
        for the measure $\widehat {H}(T)$ to be a densely defined symmetric
        operator; and the equation (2.23) implies an
        implicit form of the same assumption.

        As we will see in the next Section, the relation between
        coefficients $\{\ell_k\}$ and $\{{\cal  L}_k\}$
        can be also derived from the strong resolvent limit for a family
        of the Schr\"odinger  Hamiltonians. Since the
        conclusions of \S2 look unusual, we consider
        the weak and the resolvent limits for an explicitly
        solvable model in   Fock space. By this reason
        we restrict ourself to the case of Hamiltonians with commuting
        coefficients.

\begin{sloppypar}
\section{The strong resolvent limit of the Schr\"odinger Hamiltonians}
\end{sloppypar}
\noindent  The weak and the resolvent limits of generators of strongly
           continuous unitary groups may be different in a quite
           elementary case. The next example gives a hint to our main result.
            Consider a family of unitary groups
           $\exp\{itH_{\alpha}\}=
           U_t^{(\alpha)}:$
$$
U_t^{(\alpha)}\psi(x)=\psi(x-t)\exp\biggl\{ i\lambda\int_0^t\,d\tau\,
V_{\alpha}(x-t+\tau)\biggr\}, \quad x,\lambda\in\R,\quad \psi\in {\cal  L}_2(\R)
$$
        where $V_{\alpha}(x)=(2\pi\alpha)^{-1/2}\exp\{-x^2/2\alpha\},\;
        \alpha\in \R_+.$ Clearly $V_{\alpha}(x)\to \delta(x)$ as
        $\alpha \to +0,$ and $\int_0^t\,d\tau\,
        V_{\alpha}(x-t+\tau)\to I_{[0,t)}(x),$ with
        an indicator function $I_{\Gamma}(x)$  of the borelean set ${\Gamma}.$
        Therefore, {\it the weak limit} of the family of
        essentially self-adjoint operators
        $H_{\alpha}=i\partial_x +\lambda V_{\alpha}(x)$
        generates a  well-defined bilinear form on $W_2^1(\R)$
$$
H_*[\varphi,\psi]=(\varphi,\widehat H_w\psi)=
i(\varphi,\psi') +\lambda\overline\varphi(0)\psi(0)
$$
        that corresponds to a formal operator
         $\widehat H=i\partial_x +\lambda \delta(x).$

        On the other hand,
        {\it the strong limit} of the family of unitary groups
        $U_t^{(\alpha)}$
$$
U_t\psi(x)=e^{it\widehat {\bf H}}=\lim_{\alpha\to+0}U_t^{(\alpha)}\psi(x)=
\psi(x-t)e^{ i\lambda I_{[0,t)}(x)}=
\psi(x-t)\biggl\{(e^{i\lambda}-1) I_{[0,t)}(x)+1\biggr\}
$$
        implies {\it the strong resolvent limit} $\widehat{\bf H}$
        which can be described by a formal generator
$$
\widehat{\bf H}=r-\lim H_{\alpha}=
i\partial_x +i\bigl(e^{i\lambda}-1\bigr) \delta(x)
$$
        or by bilinear form
$$
{\bf H}_*[\varphi,\psi]=\lim_{t\to 0}{d\over dt}(\varphi,U_t\psi)=
i(\varphi,\psi')+i\bigl( e^{i\lambda}-1 \bigr)\overline\varphi(0)\psi(0).
$$
        Important observation is that $ \widehat{\bf H}=
        r-\lim H_{\alpha} \ne \widehat{H}=w-\lim H_{\alpha}.$
        The multiplier $e^{i\lambda}-1,$ an analog of
        the factor $W-I={\cal L}_3$ in equation
        (2.23), appears in a manifest form in the equation for
        $\widehat{\bf H}$. A similar phenomena happen in what follows.

        The range of the resolvent of the limit unitary group $U_t$
        is a natural domain for the generator $\widehat {\bf H}.$
        In this example it can be described explicitly by an equation:
$$
R_{\mu}\psi(x)=\int_0^{\infty} \,dt\,e^{-\mu t}\psi(x-t)+
\theta(x)\bigl( e^{i\lambda}-1 \bigr)e^{-\mu x}\int_0^{\infty}
\,dt\,e^{-\mu t}\psi(-t),
$$
	where $\theta(x)$ is an indicator function of the half--line $\R_+$.
        The structure of the resolvent shows that the functions
        from the domain of its generator
        $\widehat {\bf H}$ have  phase jumps at the origin $x=0:$
$
\lim_{x\to+0} R_{\mu}\psi(x)=e^{i\lambda}\lim_{x\to-0} R_{\mu}\psi(x).
$
        Hence the domain  ${\cal  D}_{\lambda}$
        of $\widehat {\bf H}$  consists of functions with a standard
        discontinuity at the origin:
$$
\psi: \;\psi\in W_2^1(\R\setminus\{0\}),\quad
\lim_{x\to+0} \psi(x)=e^{i\lambda}\lim_{x\to-0} \psi(x)
$$
	 and the operator $\widehat {\bf H}$ acts as $i\partial_x$ if
         $x\neq0.$
        The left and the right limits exist at the origin
        for every function from  ${\cal  D}_{\lambda}$
        because of the imbedding $W_2^1(\R\setminus\{0\})\subset C(\R\setminus\{0\}).$
        Integration by parts and the identity
$
\overline{\phi(x)}\psi(x)\bigr|_{+0}^{-0}=0
$
        for functions from ${\cal  D}_{\lambda}$
        proves that the operator $\widehat {\bf H}$ is symmetric.
        The existence of a solution from ${\cal  D}_{\lambda}$
        for the problem
        $(\widehat {\bf H}+i\mu)\psi(x)=f(x),\; x\neq 0$
        with the boundary condition as above for every $f\in{\cal  L}_2(\R),\;
        \mu=\pm 1$ implies the self-adjoint property
        of $\widehat {\bf H}.$

        Consider the Schr\"odinger equation
	${\partial_t}\psi_t=iH_{\alpha}\psi_t$
	with the self-adjoint Ha\-mil\-to\-ni\-an
        $H_{\alpha}=H\otimes I+I\otimes \widehat{\bf E}+
        H^{(\alpha)}_{int}$ depending on a scaling parameter
        $\alpha\in(0,1]:$
$$
H^{(\alpha)}_{int}= K\otimes A^+(g_{\alpha}) A(g_{\alpha})+
R\otimes A^+(f_{\alpha})+R^*\otimes A(f_{\alpha}).
\eqno(3.1)
$$
        Assume that coefficients
        $H,\,K,\,R\in {\cal  C}({\cal  H})$ have a joint
        spectral family $E_{\lambda}:$
$$
H=\int\,\nu_{\lambda}\,dE_{\lambda},\quad K=\int\,\lambda\,dE_{\lambda},\quad
R=\int\,\rho_{\lambda}e^{i\Phi_{\lambda}}\,dE_{\lambda},
$$
        where $\nu,\,\rho,\,\Phi$  are measurable real functions
        corresponding to operators $H,$ $K,$ and $R;$
        and functions $f_{\alpha}(\omega)=f(\alpha\omega),\,
        g_{\alpha}=g(\alpha\omega)$
        $f,\,g$ are elements of the space $L^+_{2,\widetilde 1}(\R)=
        \{f: \;f\in L_2(\R),\;\widetilde f=
        {\cal  F}^{-1}_{\omega\to t}f\in L_1^+(\R)\}, \; f(0)=g(0)=(2\pi)^{-1/2}.$
        We omit the indices $\alpha,$ when there is no conflicts.

        Denote by $\widehat P_t=
        \widehat P^{(\alpha)}_t(\lambda)$ a one-parameter unitary group
        in $L_2(\R)$ with the generator
$
\widehat N_{\alpha}(\lambda)=\omega+\lambda |g_{\alpha}\rangle\langle g_{\alpha}|
$
        and prove that the solution of the Schr\"odinger
        equation  $U_t^{(\alpha)}=\exp\{iH_{\alpha}t\}$ acts as follows
$$
U^{(\alpha)}_th\otimes\psi(v)=\int e^{i\nu_{\lambda} t}\,dE_{\lambda}\,
h\otimes\psi\biggl( \widehat P_t(\lambda)
v+i\rho_{\lambda} e^{i\Phi_{\lambda}}
\int_0^t \,\widehat P_s(\lambda)f_{\alpha}\,ds\biggr)
$$
$$
\times\exp\biggl\{i\rho_{\lambda} e^{-i\Phi_{\lambda}}
\int_0^t \,(f_{\alpha},\widehat P_{s}(\lambda)v)\,ds
-\rho_{\lambda}^2
\int_0^t\,dr\int_0^r\,ds\,(f_{\alpha},\widehat P_{r-s}(\lambda)
f_{\alpha}) \biggr\}.
\eqno(3.2)
$$
         For any bounded weakly measurable function
        $\widehat P_t:\R\to {\cal  B}({\cal  H})$
         the following equations are  consequences of
        linear changes of variables:
\begin{eqnarray*}
\int_0^t\,\widehat P_{r}\,dr&+&
\int_0^s\,\widehat P_{r+t}\,dr
=\int_0^{t+s}\,\widehat P_{r}\,dr,\\
\biggl(\int_0^t\int_0^r&+&\int_0^{s}\int_0^{r}\biggr)
\widehat P_{r-p}\,dr\,dp+
\int_0^{t}\int_0^s \,\widehat P_{r+p}\,dr\,dp\\
=\biggl(\int_0^t\int_0^r&+&\int_t^{t+s}\int_t^{t+r}+
\int_0^{t}\int_t^{t+s} \biggr)\,\widehat P_{r-p}\,dr\,dp=
\int_0^{t+s}\int_0^{r} \,\widehat P_{r-p}\,dr\,dp,
\end{eqnarray*}
        which are related to a group property of $U_t^{(\alpha)}:$
        the first equation generates the group property of the argument of the
        coherent vector, and the second  one generates the same property
        of the normalizing factor in (3.2).

       The weak form of the Schr\"odinger equation
       can be verified by differentiation of the quadratic form
       $\varphi_t=(h\otimes\psi(v),U^{(\alpha)}_t\,h\otimes\psi(v)):$
$$
\varphi_t=\int\,e^{i\nu_{\lambda}t}(dE_{\lambda}h,h)
\exp\biggl\{\biggl(v,\widehat P_t(\lambda)
v+i\rho_{\lambda} e^{i\Phi_{\lambda}}
\int_0^t \,\widehat P_s(\lambda)f_{\alpha}\,ds\biggr)+
$$
$$
+i\rho_{\lambda} e^{-i\Phi_{\lambda}}
\int_0^t \,(f_{\alpha},\widehat P_{s}(\lambda)v)\,ds-\rho_{\lambda}^2
\int_0^t\,dr\int_0^r\,(f_{\alpha},
\widehat P_{r-s}(\lambda)f_{\alpha})\,ds\biggr\}.
$$
        We have clearly
\begin{eqnarray*}
-i\varphi'_t\biggl|_{t=0}&=&
e^{||v||^2}\int\,(dE_{\lambda}h,h)
\biggl(\nu_{\lambda}+\lambda |(g_{\alpha},v)|^2 \\
&+&2\,{\rm Re\,}\rho_{\lambda}
e^{i\Phi_{\lambda}}(v,f_{\alpha})+\int\,d\omega\,\omega|v(\omega)|^2\biggr)
=(h\otimes\psi(v),H_{\alpha}h\otimes\psi(v)).
\end{eqnarray*}
        The one-parameter group $U^{(\alpha)}_t$ is unitary because of
        basic properties of spectral operators $E_{\lambda}$
        and the equation for the norm of an exponential vector:
$$
\biggl\Vert \psi\biggl( \widehat P_{t}v+
i\rho e^{i\Phi}
\int_0^t \,\widehat P_{s}f\,ds\biggr)
\biggr\Vert^2=
$$
$$
=\exp\biggl\{ ||v||^2+
\rho^2\biggl\Vert \int_0^t \,\widehat
P_{s}f\,ds\biggr\Vert^2-
2\rho{\,\rm Im\,} e^{-i\Phi}
\int_0^t \,(f,\widehat P_{s}v)\,ds\biggr\}.
\eqno(3.3)
$$
         Since  $\widehat P^{(\alpha)}_{t}$ is a unitary  group, we have
$$
2{\rm Re\,}\int_0^t\,dr\int_0^r\,ds\,\widehat P_{r-s}=
\overline {\int_0^t\,dr\,\widehat P_{s}}\,
\int_0^t\,ds\,\widehat P_{r}.
$$
       Hence, all terms in (3.3)  besides $ ||v||^2$ are canceled out
       by the corresponding terms in the normalizing factor.
       Thus, we obtain
\begin{theorem} The unitary group of operators
                $U_t^{(\alpha)}=\exp\{itH_{\alpha}\}$ acts as follows
$$
U_t^{(\alpha)}h\otimes\psi(v)=\int \,e^{i\nu_{\lambda} t}dE_{\lambda}\,
h\otimes\psi\biggl( \widehat P^{(\alpha)}_t(\lambda)
v+i\rho_{\lambda} e^{i\Phi_{\lambda}}
\int_0^t \,\widehat P^{(\alpha)}_s(\lambda)f_{\alpha}\,ds\biggr)\times
$$
$$
\times\exp\biggl\{i\rho_{\lambda} e^{-i\Phi_{\lambda}}
\int_0^t \,(f_{\alpha},\widehat P^{(\alpha)}_{s}(\lambda)v)\,ds
-\rho_{\lambda}^2
\int_0^t\,dr\int_0^r\,(f_{\alpha},\widehat P^{(\alpha)}_{r-s}(\lambda)
f_{\alpha})\,ds \biggr\}.
\eqno(3.4)
$$
\end{theorem}
\vskip2mm\noindent
        Equation (3.4) contains four terms depending on $\alpha$ in the right
        hand side. The next Lemma "On Four Limits" (see
         \cite{C96})
        justifies the limit of (3.4) as $\alpha\to0.$
\vskip2mm
\begin{lemma}  Let $\widehat P^{(\alpha)}_{t}(\lambda)$
              be one-parameter unitary group in  $L_2(\R)$
              generated by $\widehat N_{\alpha}(\lambda)$.
                  Then for $f,\,g\in L^+_{2,\widetilde 1}(\R)$
                  such that $f(0)=g(0)={1\over\sqrt{2\pi}}$
                  there exist limits:
\vskip1mm
\begin{itemize}
\item[\rm (1)]
$\int^t_0\,ds(g_{\alpha},\widehat P^{(\alpha)}_{s}(\lambda)
      f_{\alpha})\to (2-i\lambda)^{-1};$
\vskip1mm
\item[\rm (2)]
$\int^t_0\,ds\,\widehat P^{(\alpha)}_{s}(\lambda)
           f_{\alpha}(\omega)\to
           e^{i\omega t}\widetilde I_{(0,t)}(\omega)(1-i\lambda/2)^{-1} ; $
\vskip1mm
\item[\rm (3)]
$(g_{\alpha},\widehat P^{(\alpha)}_{t}(\lambda)v)
            \to(1-i\lambda/2)^{-1}{\cal  F}^*_{\omega\to t}v;$
\vskip1mm
\item[\rm (4)]
$ \widehat P^{(\alpha)}_{t}(\lambda)\to
          \exp\{iZ(\lambda)\widehat \pi_{(0,t)}\}=
          \widehat P_{t}(\lambda),\quad $ $\exp\{iZ(\lambda)\}=
                  (2+i\lambda)/(2-i\lambda),$
\end{itemize}
\par\noindent
            with $\widehat \pi_T$ a projector in $L_2(\R):$
                  $\widehat \pi_T={\cal  F}^*_{t\to\omega}I_T(t)
                  {\cal  F}_{\omega\to t}.$  Limits {\rm (2), (3)}
                  are weak in $L_2(\R),$ the
                  limit {\rm (4)} is strong in $L_2(\R).$
\end{lemma}
{\bf Proof.} Limits (1-3) can be derived from the Taylor decomposition
                 of the operator $\widehat P^{(\alpha)}_{t}(\lambda)$ in a
power  series $\sum C_k(\alpha,t)\lambda^k$
                  (as in \cite {AFLu, ALV}) and
                  sequential limiting procedure in each term as $\alpha\to 0.$
                  This operation maps the series into a geometrical
                  progression that converges absolutely for $|\lambda|<2.$
                The Fatou--Privalov theorem enables to extend this result
                to the entire real line  $\lambda\in \R,$ provided
                there exist a uniform  absolute bound for the series
                in the upper half-plain ${\rm\; Im} \lambda\geq 0.$ We can
                prove the required bound for the series corresponding to
                limit (4), but we do not have such bounds to prove limits (1)-(3).
                The method used below consists in application of
                properties of completely monotone functions
                (see \cite{F}, Chap. XIII, \S4) and the Duhamel equation
                 for the group $\widehat P^{(\alpha)}_{t}(\lambda):$
$$
\widehat P^{(\alpha)}_{t}(\lambda)=e^{i\omega t}+i\lambda\int_0^t
\,ds\,e^{i\omega (t-s)}|g_{\alpha}\rangle\langle g_{\alpha}|\,
\widehat P^{(\alpha)}_{s}(\lambda).
\eqno(3.5)
$$
\vskip2mm\noindent
        (1) Set  $a_t=a_t^{(\alpha)}(\lambda)=
        (g_{\alpha},\widehat P^{(\alpha)}_{t}(\lambda)f_{\alpha}).$
        As a consequence of (3.5), we have
$$
\int_0^t \,a_s\,ds=\Phi_t(g_{\alpha},f_{\alpha})+i\lambda\int_0^t\,
\Phi_{t-s}(g_{\alpha},g_{\alpha}) \,a_s\,ds,
\eqno(3.6)
$$
        with $\Phi_t(g,f)=\int_0^t\,ds\,
        (g,e^{i\omega s}f)=\mu_{f,g}(0,t),$ where
        $\mu_{f,g}(\cdot)$ is2 the probability measure such that
$$
\mu_{f,g}(T)=\int_T\, ds\int_{\r}\,dt\,
\overline {\widetilde g(s-t)}\widetilde f(t),\quad \mu_{f,g}(\R)=2\pi g(0)f(0)=1.
$$
        Denote by
        $\widetilde a_p=\widetilde a_p^{(\alpha)}(\lambda)$ the Laplace
        transform of the bounded continuous function
        $a_t.$  Since the equation (3.6) contains a convolution, we obtain an
        equivalent algebraic  equation
$
p^{-1}\widetilde a_p= \widetilde \Phi_p(g,f) +i\lambda \widetilde a_p
\widetilde \Phi_p(g,g).
$
        The solution reads as
$$
\widetilde a_p=p\widetilde \Phi_p(g,f)
(1-i\lambda p\widetilde \Phi_p(g,g))^{-1},
$$
        where $p\widetilde \Phi_p(g,f)=\int_0^{\infty}\,e^{-pt}
        d\,(g,e^{i\omega t}f)=
        \widetilde\mu_{f,g}(p)$ is the Laplace transform of the probability
        measure $\mu_{f,g}$.
        Since $\Phi_t(g_{\alpha},f_{\alpha})=\mu_{f,g}(0,t/{\alpha}),$
        then $\widetilde a^{(\alpha)}_p(\lambda)
        =\widetilde a_{p\alpha}(\lambda)$ and
$$
\widetilde a_p(\lambda)=
{\widetilde\mu_{f,g}( p)
\over1-i\lambda\widetilde\mu_{g,g}( p)}=
{\widetilde\mu_{f,g}( p)
\over1+|\lambda|^2\widetilde\mu^2_{g,g}( p)}+i\overline\lambda
{\widetilde\mu_{f,g}(p)\widetilde\mu_{g,g}( p)
\over1+|\lambda|^2\widetilde\mu^2_{g,g}(p)}.
$$
        We recall that the class of the Laplace transforms of
         positive measures coincides with the class of completely
        positive functions. It contains the function
        $(1+|\lambda|^2p)^{-1},$  and it is closed w.r.t.
        addition, multiplication and composition
        (see. \cite{F}, Chap. XIII, \S4).
        Hence, the real and imaginary parts of
        $\widetilde a_{p\alpha}(\lambda)$ are the Laplace transforms
        of bounded measures:
$$
\widetilde a^{(\alpha)}_p(\lambda)=\int_0^{\infty} \,e^{-pt}
\biggl( A_{\lambda}(\alpha^{-1}dt)+i\lambda B_{\lambda}(\alpha^{-1}dt)\biggr),
$$
$$
A_{\lambda}(\R_+)=
{\mu_{f,g}(\R_+)
\over1+|\lambda|^2\mu^2_{g,g}(\R_+)}={2\over4+\lambda^2},
\quad
B_{\lambda}(\R_+)={\mu_{f,g}(\R_+)\mu_{g,g}(\R_+)
\over1+|\lambda|^2\mu^2_{g,g}(\R_+)}={1\over4+\lambda^2}
$$
        It follows from here  for any $\lambda \in \C$ that
$$
\int_T \,a_s^{(\alpha)}(\lambda)\,ds=
A_{\lambda}(T/\alpha)+i\lambda B_{\lambda}(T/\alpha)\qquad \forall
T\in {\cal  B}(\R_+)
\eqno(3.7)
$$
$$
\lim_{\alpha\to+0}\int_0^t \,a_s^{(\alpha)}(\lambda)\,ds=
A_{\lambda}(\R_+)+i\lambda B_{\lambda}(\R_+)=(2-i\lambda)^{-1}.
$$

\vskip3mm
\par\noindent
             (2)  Denote
             $\beta_t^{(\alpha)}(\lambda,\omega)=\int_0^t\,ds\,
             \widehat P^{(\alpha)}_{s}(\lambda)f_{\alpha}(\omega),$
             and for  $v\in L_2(\R)$ set
$$
b_t^{(\alpha)}(\lambda,v)=\int\,d\omega\,\overline v(\omega)
\beta_t^{(\alpha)}(\lambda,\omega),
\quad
\varphi_t(v,f)=\int_0^t\,ds\,(v,e^{i\omega s}f).
$$
           From equations (3.5) and (3.7) follows an identity
$$
b_t^{(\alpha)}=\varphi_t(v,f_{\alpha})+
i\lambda\int_0^t\,ds\,\varphi_{t-s}(v,g_{\alpha})\int_0^{t-s}\,d\tau\,
(g_{\alpha},\widehat P^{(\alpha)}_{\tau}(\lambda)f_{\alpha})=
$$
$$
=\varphi_t(v,f_{\alpha})+
i\lambda\int_0^t\,ds\,\varphi_{t-s}(v,g_{\alpha})
\biggl( A_{\lambda}(\alpha^{-1}(t-s))+
i\lambda B_{\lambda}(\alpha^{-1}(t-s)) \biggr).
\eqno(3.8)
$$
        The integral $\int_T \,dt e^{i\omega t}g_{\alpha}(\omega)$
        converges in $L_2(\R)$ to
        $ e^{i\omega t}\widetilde I_{(0,t)}(\omega)$
        as $\alpha \to +0.$ Since  $L_2(\R)\in L_1^{loc}(\R),$ the functions
        $\varphi_t(v,f_{\alpha})$ and $\varphi_t(v,g_{\alpha})$ converge
        to $(v,e^{i\omega t}\widetilde I_{(0,t)}).$
        The function $A_{\lambda}(\alpha^{-1}t)+i\lambda
        B_{\lambda}(\alpha^{-1}t)$
        is measurable in $t$, bounded uniformly in $\alpha\in (0,1]$
        and converges at each point to
        $(2-i\lambda)^{-1}$ as $\alpha \to 0.$
        Therefore, it is possible to pass to the limit in (3.8):
$
b_p^{(\alpha)}(\lambda,v)
\to\bigl( 1+i\lambda/(2-i\lambda) \bigr)
(v,e^{i\omega t}\widetilde I_{(0,t)}).
$
        Hence, for every $v\in L_2(\R)$
$$
\lim_{\alpha\to0}
\int^t_0\,ds\,(v,\widehat P^{(\alpha)}_{s}(\lambda)f_{\alpha})=
(1-i\lambda/2)^{-1}(v,e^{i\omega t}\widetilde I_{(0,t)}),
\eqno(3.9)
$$
       that is $w-\lim \beta_t^{(\alpha)}(\lambda,\omega)=
       (1-i\lambda/2)^{-1}e^{i\omega t}\widetilde I_{(0,t)}(\omega).$
\vskip3mm\noindent
                  (3) Rewrite (3.9) in an equivalent form:
$$
\lim_{\alpha\to0}
\int^t_0\,ds\,(v,\widehat P^{(\alpha)}_{s}(\lambda)f_{\alpha})=
{2\over2-i\lambda}\int\,\overline {{\cal  F}_{\omega\to \tau}v}\,
I_{(0,t)}(\tau)\,d\tau.
\eqno(3.10)
$$
       Since $(g_{\alpha},P^{(\alpha)}_t(\lambda)v)=
       \overline {(v,P^{(\alpha)}_{-t}(\lambda)g_{\alpha})},$
       from  (3.10) follows an equation
$$
\lim_{\alpha\to0}
\int^t_0\,ds\,(g_{\alpha},\widehat P^{(\alpha)}_{s}(\lambda)v)=
\overline {{2\over2+i\lambda}({\cal  F}^*_{\omega\to \tau}v,I_{(0,t)}(\tau))}=
{2\over2-i\lambda}(I_{(0,t)}(\tau),{\cal  F}^*_{\omega\to \tau}v).
\eqno(3.11)
$$
        The set of the indicator functions of bounded Borelean subsets
        in $\R$ is total in $L_2(\R).$ Therefore, from  (3.11) follows
$$
\lim_{\alpha\to0}
\int\,ds\,\overline r(t)(g_{\alpha},\widehat P^{(\alpha)}_{s}(\lambda)v)=
{2\over2-i\lambda}( r(\cdot),{\cal  F}^*_{\omega\to \cdot}v)
$$
                for every $r,v\in L_2(\R),$
        or $w-\lim\, (g_{\alpha},\widehat P^{(\alpha)}_{t}(\lambda)v)=
        (1-i\lambda/2)^{-1}{\cal  F}^*_{\omega\to t}v$ in $L_2(\R).$
        From here we conclude that the norm of the weakly converging
        family
$$
g_{\alpha,t}=(g_{\alpha},\widehat P^{(\alpha)}_{t}(\lambda)v)\in L_2(\R)
$$
        is  uniformly bounded for every  $\alpha\in(0,1].$
\vskip3mm\noindent
           (4) Note that the strong convergence of
           unitary operators follows from the weak convergence, and the weak convergence
           follows from the convergence ot corresponding quadratic forms.
           Hence, it suffices to prove  that
           $(v,\widehat P_t^{(\alpha)}(\lambda)v)\to
           (v,\widehat  P_t(\lambda)v)$  $\forall v\in L_2(\R).$
           Denote $c_t^{(\alpha)}=c_t^{(\alpha)}(\lambda,v)=
           (v,\widehat  P_t^{(\alpha)}(\lambda)v).$
           From (3.1) follows the identity:
$$
c_t^{(\alpha)}=(v,e^{i\omega t}v)+i\lambda
\int_0^t\,ds \,(v,e^{i\omega (t-s)}g_{\alpha})
(g_{\alpha},\widehat  P^{(\alpha)}_s(\lambda)v),
\eqno(3.12)
$$
        where the sequence $g_{\alpha,s}=
        (g_{\alpha},\widehat  P^{(\alpha)}_s(\lambda)v)$
        is uniformly bounded and weakly converging in $L_2(\R).$
        Set $\widetilde v(t)={\cal  F}_{\omega\to t}v(\omega),$ and
        $\widetilde v_{\alpha}(t)=\overline{(v,e^{i\omega t}g_{\alpha})}.$
        Since the function $g(\omega)={\cal  F}_{t\to\omega}\widetilde g(t)$
         can be decomposed into an integral,
        the square of the norm of $\delta_{\alpha}=
        ||\widetilde v-\widetilde v_{\alpha}||$
        can be represented as an integral of absolutely integrable functions
$$
\delta_{\alpha}^{2}=\int_{\R^3}\,dt\,dr\,ds\,
(\overline{\widetilde v(t-\alpha s)}-\overline{\widetilde v(t)})\,
\overline{\widetilde g(s)}\,
{(\widetilde v(t-\alpha r)-\widetilde v(t))\,\widetilde g(r)}.
$$
        Clearly, $\delta_{\alpha}\to0$ because
        it is possible to pass to the limit  in this integral as
        $\alpha\to0$ for every $v\in L_2(\R),\;
        g\in L^+_{2,\widetilde 1}(\R).$ Therefore, we prove the norm
        convergence  $\widetilde v_{\alpha}\to
        \widetilde v.$  Now we can apply the equation (3) to the
        weakly converging
        sequence $(g_{\alpha},\widetilde P^{(\alpha)}_s(\lambda)v)$
        and pass   to the limit in the integral (3.12):
$$
c_t^{(\alpha)}\to
\biggl( v, e^{i\omega t}\bigl\{I+ {2i\lambda\over2-i\lambda}
\int_0^t\,ds\,e^{-i\omega s}{\cal  F}^*_{\omega\to s} \bigr\}v \biggr)=
\biggl( v, e^{i\omega t}\bigl
\{I+ {2i\lambda\over2-i\lambda}\widehat \pi_{(0,t)} \bigr\}v \biggr).
$$
        In this way we obtain the weak and the strong limits
$$
\lim_{\alpha\to0}\widehat P_t^{(\alpha)}(\lambda)=
I+ {2i\lambda\over2-i\lambda}\widehat \pi_{(0,t)}=
{2+i\lambda\over2-i\lambda}\widehat \pi_{(0,t)}+\widehat \pi_{(0,t)}^{a}=
\exp\{iZ(\lambda)\widehat \pi_{(0,t)}\},
$$
        with the projector-valued family  $\widehat \pi_T$
        from ${\cal  B}(L_2(\R)),$
        and $I-\widehat \pi_T=\widehat \pi_{T^a}.$

        \qed

        Limits (1)-(4) describe what occur with the
        solution (3.3) as $\alpha\to0.$
        Substituting (1)-(4) to (3.3) we obtain the unitary group
        $U_t=\exp\{i\widehat {\bf H}t\}=s-\lim\,U_t^{(\alpha)}$:
\begin{eqnarray*}
U_t\,h\otimes \psi(v)&=&
\int\,e^{-G_{\lambda}t}\,dE_{\lambda}\,h\otimes
\psi\biggl(e^{iZ(\lambda)\widehat\pi_{(0,t)}}e^{i\omega t}v+
i\rho_{\lambda}e^{i\Phi_{\lambda}}
{2\over2-i\lambda}\widetilde I_{(0,t)}  \biggr)\\
&\times&\exp\biggl\{i\rho_{\lambda}e^{-i\Phi_{\lambda}}
{2\over2-i\lambda}(\widetilde I_{(0,t)},e^{i\omega t}v) \biggr\},
\qquad \qquad \qquad \qquad \qquad (3.13)
\end{eqnarray*}
        with $G_{\lambda}=-i\nu_{\lambda}-
        \rho^2_{\lambda}/(2-i\lambda).$
        The group property of $U_t$ follows from (3.13) and from
        the properties of the Fourier transform:
$$
e^{i\omega t} \widehat\pi_T e^{-i\omega t}=\widehat\pi_{T+t},\quad
e^{i\omega t}\widetilde I_{(0,t)}(\omega)=
\overline{\widetilde I}_{(0,t)}(\omega).
$$
        Taking the time derivative
$$
\lim_{t\to+0}
{1\over i}{d\over dt}(g\otimes\psi(f),U_t\,h\otimes\psi(v))
={\bf H}_*[g\otimes\psi(f),h\otimes\psi(v)]
$$
  where $g,h\in D\subseteq{\cal  H}, \quad f,v\in \widetilde W_2^1(\R)$
       we obtain a bilinear form
$$
{\bf H}_*[g\otimes\psi(f),h\otimes\psi(v)]=
e^{(f,v)_{L_2}}\biggl((g,H_0h)_{\cal  H}+
(g,H_1h)_{\cal  H}\,{\widetilde v(0)}
+(g,H_2h)_{\cal  H}\,\overline{\widetilde f(0)}
$$
$$
+(g,h)_{\cal  H}\,\int\, d\omega \overline f(\omega)g(\omega)\,\omega+
\bigl(g,H_3h\bigr)_{\cal  H}
\,\overline{\widetilde f(0)}\widetilde v(0)\biggr),
$$
   where
$$
{ H}_0=H-R^* {K\over {4+K^2}}R+
R^*{2i\over{4+K^2}}R=iG,\quad { H}_1=R^*{2\over{ 2-iK}},
$$
$$
{ H}_2={2\over{ 2-iK}}R,\quad
{ H}_3={2K\over{2-iK}}=i(I-W),\quad  W={2+iK\over2-iK}.
\eqno(3.14)
$$
        In what follows we assume that the operator
        $G =-iH+{i\over4}L^*KL+{1\over2}L^*L$ is a
        generator of one-paremeter contraction
        semigroup  $W_t=\exp\{-Gt\}$
        in $\cal H$ such that
$$
D=dom\, H\cap dom\,L^*L\subseteq dom\,G\subseteq dom\,L,\quad
G^*\phi+G\phi=L^*L\phi \quad \forall \phi \in D,
$$
        $H_s=-H+\frac{1}{4}L^*KL$ is a  operator symmetric
        on $D,$ and $D$ is core for $G$ (see \cite {CF}).

        The formal operator expression describing the quadratic form
        $ {\bf  H}_*$ reads as
        $\widehat {\bf  H}=I\otimes \widehat {E}+
        { H}_0\otimes I+(2\pi)^{-1/2}{ H}_1\otimes A(1)
        + (2\pi)^{-1/2}{H}_2\otimes A^+(1)
        +(2\pi)^{-1}{H}_3\otimes A^+(1)A(1).$

        Consider the function of a set $u(s,t)=J_sU_{t-s}J_t^*.$
        Since $J^*_t\psi(v)=\psi(e^{-i\omega t}v),$ we obtain from
        (3.13)
$$
u(T)\,h\otimes \psi(v)=
\int\,e^{-G_{\lambda}{\rm mes\,}T}\,dE_{\lambda}\,h\otimes\psi\biggl(
e^{iZ(\lambda)\widehat\pi_{T}}v+
$$
$$
+i\rho_{\lambda}e^{i\Phi_{\lambda}}
{2\over2-i\lambda}\widetilde I_T  \biggr)
\exp\biggl\{i\rho_{\lambda}e^{-i\Phi_{\lambda}}
{2\over2-i\lambda}(\widetilde I_T,v) \biggr\}.
$$
        The $u(T)$ family of operators is interval adapted,
        commutative for disjoint arguments, and satisfies
        the cocycle composition rule $u(T_1\cup T_2)=u(T_1)u(T_2),\quad
        T_1\cap T_2=\emptyset.$
        The weak evolution equation for $u(T)$ reads
$$
d(h\otimes\psi(v),u(0,t)h\otimes\psi(v))=
i(h\otimes\psi(v),u(0,t)H(dt_+)h\otimes\psi(v)),
$$
        where $i\widehat H(T)=M(T)=i\int_T\,dt( J_t\widehat{\bf H}J^*_t
        -\widehat {\bf E}\otimes I)$ is well-defined operator-valued measure:
\begin{eqnarray*}
\widehat H(T)&=&\biggl( H-R^*{K\over4+K^2}R +R^*
{2i\over4+K^2}R \biggr)\otimes {\rm mes\,}T\\
&+&{2\over2-iK}R\otimes A^+(T)+
R^*{2\over2-iK}\otimes A(T)+i(I-W)\otimes \Lambda(T).
\end{eqnarray*}
        Hence, we obtain the following result.
\begin{theorem} The family of solutions  for the Schr\"odinger equation
             with Hamiltonian (3.1) converges in $L_2(\R)$ to the solution
             of stochastic differential equation
             (2.10) with coefficients (2.23): $u(0,t)=s-\lim_{\alpha\to 0}
             U^{(\alpha)}_tJ^*_t.$ Conditions (3.14) are necessary for
             the generator $\widehat{\bf H}$
              of the limit unitary group $U_t$ to be symmetric.
\end{theorem}

\bigskip
\begin{sloppypar}
\section{Surprises of the resolvent}
\end{sloppypar}
        Consider a Fock vector
        $\Phi \in{\bf h}= {\cal  H}\otimes \Gamma^S({\cal  L}_2(\R))$
        belonging to the range of the resolvent
$$
\Phi=R_{\mu} h\otimes\psi(v)=\int_0^{\infty} \,dt\,e^{-\mu t}
U_t \,h\otimes\psi(v)=\{\Phi_n(\omega)\},\quad
$$
$$
\Phi_n(\cdot):\R^n\to{\cal  H},\quad \omega=\{\omega_1,\dots, \omega_n\}
$$
        with components (3.13):
\begin{eqnarray*}
\Phi_n(\omega)&=&\int_0^{\infty} \,dt\,
\exp\biggl\{-\mu(G+ t)-L^*W\int_0^t\,\widetilde v(-\tau)\,d\tau\biggr\}
        \phi_{n,t}(\omega),\\
\phi_{n,t}(\omega)
&=&\prod_1^n \biggl((W-1)\pi_{[0,t)}e^{i\omega_k t}v(\omega_k)
+e^{i\omega_k t}v(\omega_k)+L\widetilde I_{[0,t)}(\omega_k) \biggr)h,
\end{eqnarray*}
        for commuting operators $L,\, W,$ and $ G$ as above:
$$
L=\int\,dE_{\lambda}\,L(\lambda),\quad
W=\int\,dE_{\lambda}\,W(\lambda),\quad
G=\int\,dE_{\lambda}\,G(\lambda),
$$
$$
L(\lambda)=2i\rho(\lambda)e^{-i\Phi(\lambda)}(2-i\lambda)^{-1},\quad
W(\lambda)=e^{iZ(\lambda)},\quad G(\lambda)=-i\nu(\lambda)+
\rho(\lambda)^2/(2-i\lambda).
$$
        Denote by $\widetilde\phi_{n,t}$
        the Fourier transform
of the function $\phi_{n,t}(\omega)$  with respect to variables
$\omega=\{\omega_1,\dots,\omega_n\}:$
$$
\widetilde\phi_{n,t}(\tau)=
\prod_1^n \biggl((W-I)I_{[0,t)}(\tau_k)
\widetilde {v}(\tau_k-t)
+\widetilde {v}(\tau_k-t)+L\,I_{[0,t)}(\tau_k) \biggr)h,
$$
where $\tau=\{\tau_1,\dots,\tau_n\}.$
        Let  ${\cal  K}$ be a subset in $\{1,\dots,n\}$ and let
        ${\cal  K}^a$  be its complement. Put
$$
P_{{\cal  K},t}^{(n)}(\tau)=
\prod_{k\in{\cal  K}} \biggl((W-I)\widetilde {v}(\tau_k-t)
+L \biggr)I_{[0,t)}(\tau_k)\in {\cal  B}(\bf h).
$$
        then
$$
\widetilde\phi_{n,t}(\tau)=\sum_{{\cal  K}}
\biggl( P_{{\cal  K},t}^{(n)}(\tau) \,\prod_{m\in{\cal  K}^a}
\widetilde {v}(\tau_m-t)\biggr)h.
\eqno(4.1)
$$
             The functions
        $P_{{\cal  K},t}^{(n)}(\tau)$  have  discontinuity in hyperplanes
        where variables $\tau_k$ change the sign:
$$
\lim_{\tau_k\to-0} P_{{\cal  K},t}^{(n)}(\tau)=
I_{{\cal  K}^a}(k)P_{{\cal  K},t}^{(n)}(\tau),
$$
$$
\lim_{\tau_k\to+0} P_{{\cal  K},t}^{(n)}(\tau)=
I_{{\cal  K}^a}(k)P_{{\cal  K},t}^{(n)}(\tau)
+\biggl( (W-I)\widetilde v(-t)+L \biggr)P_{{\cal  K}
\setminus\{k\},t}^{(n-1)}(\tau).
\eqno(4.2)
$$
        Therefore, from (4.2) follows:
$$
P_{{\cal  K},t}^{(n)}(\tau)\bigr|_{\tau_k=+0}^{\tau_k=-0}
=-\biggl( (W-I)\widetilde v(-t)+L \biggr)
P_{{\cal  K}\setminus\{k\},t}^{(n-1)}(\tau)\,I_{{\cal  K}}(k).
\eqno(4.3)
$$
        Let us find values of jumps of $\widetilde\phi_{n,t}(\tau)$ when
        $\tau_k$ changes the sign. Note that
$$
\lim_{\tau_k\to-0} \widetilde\phi_{n,t}(\tau)=\widetilde v(-t)
\widetilde\phi_{n-1,t}(\tau_1,\dots,\tau_{k-1},\tau_{k+1},\dots,\tau_{n}).
$$
        Taking into consideration equations (4.1) and (4.3) we obtain
        the amplitude and phase jumps for functions from the range of
        the resolvent  of the unitary group $U_t:$
$$
\lim_{\tau_k\to+0} \widetilde\phi_{n,t}(\tau)=
W \lim_{\tau_k\to-0} \widetilde\phi_{n,t}(\tau)+L\,
\widetilde\phi_{n-1,t}(\tau_1,\dots,\tau_{k-1},\tau_{k+1},\dots,\tau_{n}).
\eqno(4.4)
$$
        Denote by ${\cal  D}_{W,L}=
        D\otimes \Gamma^S(\widetilde W_2^1(\R\setminus\{0\}))$
        a vector subspace in ${\bf h}$ that satisfies the condition
        (4.4). By
        $A(\delta_{\pm}),\,\Lambda(\delta_{\pm}),$ and
        $ \widehat N$ we denote
        operators acting on symmetric Fock vectors as follows:
$$
(\Phi, \Lambda(\delta_{\pm})\Psi)=\lim _{\varepsilon\to\pm0}
\sum_1^{\infty} \frac{1}{n!}\sum_{k=1}^n
\int_{\bigl(\R\setminus\{0\}\bigr)^{n-1}}
$$
$$
\times\prod_{m\neq k}d\tau_m
\bigl(\widetilde{\Phi}_n,
\widetilde{\Psi}_n\bigr)_{\cal  H}
(\tau_1,\dots,\tau_{k-1},\epsilon,\tau_{k},\dots,\tau_{n-1}),
$$
$$
A(\delta_{\pm}\widetilde\Psi\bigr)_n(\tau)
=\lim _{\varepsilon\to\pm0}
\sum_{k=1}^n \widetilde{\Psi}_{n+1}
(\tau_1,\dots,\tau_{k-1},\epsilon,\tau_{k},\dots,\tau_{n}),\quad
\widehat N\Psi_n(\omega)=n\Psi_n(\omega).
$$
         The boundary condition (4.4) in this notation looks as follows
$$
(\widehat N+1)^{-1}(I\otimes A(\delta_+)-W\otimes A(\delta_-))\Psi=
(L\otimes I)\,\Psi.
\eqno(4.5)
$$

        Let us prove that the operator
$$
\widehat {\bf H}=iG\otimes I+I\otimes
\widehat {\bf E}+iL^*W\otimes A(\delta_-),\quad
\widehat {\bf E}={\cal  F}^{*}_{ \tau\to\omega}
        \int_{\R\setminus\{0\}}d\tau a^+(\tau)a(\tau)\,i\partial_{\tau}
        {\cal  F}_{\omega\to \tau},
\eqno(4.6)
$$
        is symmetric in ${\cal  D}_{W,L}$. Let
        $\Phi,\Psi\in {\cal  D}_{W,L}$ and let $B$ be a Hermitian operator so that
        $dom\, B\otimes I\supseteq{\cal  D}_{W,L} .$ Integration by parts
        implies an identity, where the difference of substitutions
        is expressed through operators $\Lambda(\delta_{\pm}):$
$$
(\Phi,B\otimes \widehat {\bf E} \,\Psi)-
(B\otimes \widehat {\bf E} \,\Phi,\Psi)=
i\biggl(\Phi,B\otimes\bigl(\Lambda(\delta_-)-
\Lambda(\delta_+)\bigr)\Psi\biggr)
\eqno(4.7)
$$
         Using boundary condition (4.4) for functions
         $\widetilde \Phi_n$ and $\widetilde \Psi_n$
         we find the difference between the substitutions (4.6):
$$
i\biggl(\Phi,B\otimes\bigl(\Lambda(\delta_+)-
\Lambda(\delta_-)\bigr)\Psi\biggr)=i(\Phi,(W^*BW-B)\Lambda(\delta_-)\Psi)
$$
$$
+i(L\,\Phi,BL\,\Psi)+
i(W\, A(\delta_-)\,\Phi,BL\, \Psi)+i(L\, \Phi,BW\,A(\delta_-)\, \Psi).
\eqno(4.8)
$$
        In the particular case $B=I,$ the equation (4.8) becomes simpler:
$$
i\biggl(\Phi,\bigl(\Lambda(\delta_+)-
\Lambda(\delta_-)\bigr)\Psi\biggr)
=i(\Phi,L^*L\Psi)
-(iL^*W\otimes A(\delta_-)\Phi,\Psi)
+(\Phi,iL^*W\otimes A(\delta_-)\Psi).
$$
	Now the identity $iG-iL^*L=(iG)^*$ and equation (4.7) prove
        that the operator $\widehat {\bf H}$ is symmetric on
        ${\cal  D}_{W,L}:$
$$
(\Phi,\widehat {\bf H}\Psi)=\bigl((I\otimes \widehat {\bf E})\,\Phi, \Psi  \bigr)+
\bigl(\Phi, \bigl\{iG\otimes I+ iL^*W\otimes A(\delta_-)\bigr\}\Psi  \bigr)
$$
$$
-i\bigl(\Phi,I\otimes\bigl(\Lambda(\delta_+)-
\Lambda(\delta_-)\bigr)\Psi\bigr)
$$
$$
=\bigl((I\otimes \widehat {\bf E})\,\Phi, \Psi  \bigr)+
\bigl(\Phi, (iG)^*\otimes I\Psi  \bigr)+
(iL^*W\otimes A(\delta_-)\Phi,\Psi))=(\widehat {\bf H}\Phi,\Psi).
$$

        Let us find how the generator of the unitary group $U_t$
        acts on Fock vectors belonging to the
        range of the resolvent. Let $\Psi\in {\bf h},\;
        \Phi=R_{\mu}h\otimes \psi(v).$
        From the definition of the generator we have:
$$
(\Psi, \widehat {\bf H}\, \Phi)=\lim_{s\to+0}
\frac{1}{i}\frac{d}{ds}(\Psi,U_s\Phi)
=\frac{1}{i}\int_0^{\infty} dt\,\sum_{n=0}^{\infty} \frac{1}{n!}
\int_{\bigl(\R\setminus\{0\}\bigr)^n}\,d\tau
$$
$$
\times
\biggl( \widetilde \psi_n(\tau),\frac{d}{ds}
e^{-(G+\mu)t-Gs-iL^*W\int_0^{t+s}
\widetilde v(-\tau)\,d\tau}
\widetilde \phi_{n,t+s}(\tau) \biggr)_{{\cal  H}}\biggr|_{s=0}.
\eqno(4.9)
$$
        Note that the functions $\widetilde \phi_{n,t}(\tau)$ depend on
        differences $\tau_k-t.$  Hence,
$$
\frac{d}{dt}\widetilde \phi_{n,t}(\tau)=
-\sum_{k=1}^n \frac{\partial }{\partial \tau_k}\widetilde \phi_{n,t}(\tau)=
i{\cal  F}_{\omega\to\tau}\widehat {\bf E}\, \widetilde \phi_{n,t}(\tau).
\eqno(4.10)
$$
        On the other hand, from the definition of the operator
        $A(\delta_-)$ we have
$$
{\cal  F}_{\omega\to \tau}\bigl({A(\delta_-)\phi_{n,t}}\bigr)
(\tau)=n\widetilde v(-t)\,\phi_{n-1,t}(\tau).
\eqno(4.11)
$$
        Now from the (4.9) and form  equations (4.10), (4.11) we obtain
$$
(\Psi, \widehat {\bf H}\, \Phi)_{\bf h}=
\int_0^{\infty} dt\,
\bigl(\psi_0,iGe^{-(G+\mu)t-iL^*W\int_0^{t}
\widetilde v(-\tau)\,d\tau}  h \bigr)_{{\cal  H}}
+\int_0^{\infty} dt\,\sum_{n=1}^{\infty} \frac{1}{n!}
\int_{\bigl(\R\setminus\{0\}\bigr)^n}\,d\tau
$$
$$
\times\biggl( \widetilde \psi_n(\tau),
e^{-(G+\mu)t-iL^*W\int_0^{t}
\widetilde v(-\tau)\,d\tau}
\bigl(iG+iL^*W\widetilde v(-t)+
i\sum_{k=1}^n \frac{\partial }{\partial \tau_k}\bigr)
\widetilde \phi_{n,t}(\tau) \biggr)_{{\cal  H}}
$$
$$
=\biggl(\Psi,\biggl\{iG+iL^*W\otimes A(\delta_-)+I\otimes\widehat {\bf E}\biggr\}
\Phi\biggr)_{\bf h},
$$
        that is the generator  $\widehat {\bf H}$
	of the group $U_t$ coincides with the generator (4.5). Thus
        we have proved the theorem.

\begin{theorem}
 The  generator
${\widehat {\bf H}}=iG\otimes I+I\otimes\widehat {\bf E}+
iL^*W\otimes  A(\delta_-)$ of one-parameter unitary group
$U_t $ is symmetric in ${\cal  D}_{W,L}.$
\end{theorem}

        We have  proved that the operator ${\widehat {\bf H}}$ is symmetric
        without the assumption that operators $L,\,G,$ and $W$ commute.
        There is no conceptual difficulties to extend the construction
        of a symmetrical boundary value problem
        to a wider class of generators
$$
 {\widehat {\bf H}}=iG+I\otimes\widehat {\bf E}+
 i\sum\limits_{\ell,m} L^*_{\ell}W_{\ell,m} \otimes A_m(\delta_-)
\eqno(4.12)
$$
         with the boundary condition
$$
(\widehat N+1)^{-1}\bigl(I\otimes A_{\ell}(\delta_+)
-\sum\limits_{m} W_{\ell,m} \otimes A_m(\delta_-)
\bigr)\Psi=(L_{\ell}\otimes I)\,\Psi,
\eqno(4.13)
$$
        where $W=\{W_{\ell, m}\}$ is $(M\times M)$ is a unitary matrices with
        coefficients from ${\cal  B}({\cal  H}),\;$
        $\{A_{\ell}(g):\;g\in L_2(\R),\,1\le\ell\le M\}$
        are annihilation operators
        $\Gamma^S(L_2(\R^M))$ which commute for different indices $\ell,$
        and $G=-iH_0+{1\over2}\sum L_{\ell}^*L_{\ell}.$

\begin{sloppypar}
\section{The Markow evolution equation}
\end{sloppypar}
         Let us consider how the Markow evolution equation can be derived
         from the boundary value problem for the Schr\"odinger equation
$$
\frac{d}{dt}\Psi(t)= \biggl(-G+iI\otimes\widehat {\bf E}
 -\sum\limits_{\ell,m} L^*_{\ell}W_{\ell,m} \otimes A_m(\delta_-) \biggr)
 \Psi(t)
$$
        with the boundary condition (4.13). Let $B$ be a Hermitian
        operator from  ${\cal  B}({\cal H})$ and let $h,\,g\in D.$
        Consider an equation for mean values
$
(g,P_t(B)h)_{{\cal  H}}=
(U_t\,g\otimes\Psi(0)|B\otimes I|U_t\,h\otimes\Psi(0))_{\bf h}.
$
        From (4.12) we have
$$
\frac{d}{dt}(g,P_t(B)h)_{{\cal  H}}=
-((G+\sum_{\ell,m}L^*_{\ell}W_{\ell,m}\, A_m(\delta_-))U_t\,
g\otimes\Psi(0)|B\otimes I|U_t\,h\otimes\Psi(0))
$$
$$
-(U_t\,g\otimes\Psi(0)|B\otimes I|
(G+\sum_{\ell,m}L^*_{\ell}W_{\ell,m}\, A_m(\delta_-))
U_t\,h\otimes\Psi(0))
$$
$$
+\biggl(U_t\,g\otimes\Psi(0)|B\otimes\bigl(\Lambda(\delta_+)-
\Lambda(\delta_-)\bigr)|U_t\,h\otimes\Psi(0))\biggr).
\eqno(4.14)
$$
        Now the equation (4.8) reads as
$$
\biggl(\Phi,B\otimes\bigl((\Lambda(\delta_+)-
\Lambda(\delta_-)\bigr)\Psi\biggr)=
\sum_{\ell,m}\biggl(\Phi,\bigl(W^*_{\ell,m}BW_{l,m}-B\bigr)
\Lambda_m(\delta_-)\Psi\biggr)
$$
$$
+\sum_{\ell}(L_{\ell}\,\Phi,BL_{\ell}\,\Psi)
+\sum_{\ell,m}\biggl(
(W_{\ell,m}\, A_m(\delta_-)\,\Phi,BL_{\ell}\, \Psi)+
(L_{\ell}\, \Phi,BW_{\ell,m}\,A_m(\delta_-)\, \Psi)\biggr)
\eqno(4.15)
$$
        Since $A(\delta_-)\Psi(0)=0,\,\Lambda(\delta_-)\Psi(0)=0,$
        (4.14) and (4.15) imply:
$$
\frac{d}{dt}(g,P_t(B)h)_{{\cal  H}}\biggr|_{t=0}=(g,{\cal  L}(B)h)=
-(Gg,Bh)-(g,BGh)+ \sum_{\ell} (L_{\ell}g,BL_{\ell}g).
$$
        Hence, we obtain an infinitesimal map  ${\cal  L}(\cdot)$  of the Markov evolution
        equation in the standard Lindbladien form:
$$
\frac{d}{dt}P_t(B)={\cal  L}(P_t(B)),\quad {\cal  L}(B)=
-G^*B-BG+\frac{1}{2}\sum_{\ell} L^*_{\ell}BL_{\ell}
$$
  for $G=-iH_0+{1\over2}\sum L_{\ell}^*L_{\ell}.$

\begin{sloppypar}
\section{Concluding remarks}
\end{sloppypar}

\noindent
         Among the most important problems which stay unsolved in this paper
         we can mention

         -a proof of {\sl srs}-convergence $U_t^{(\alpha)}J^*_t\to u(0,t)$
         for Hamiltonians with noncommuting coefficients $H,\,R,$ and $K;$

         -an extension of a theory to the case $\widehat {\bf E}=
         \int\,|\omega|^2 a^+(\omega)a(\omega)d\omega$ or
         $\widehat {\bf E}=\int\,(\omega|^2+c^2)^{1/2}
         a^+(\omega)a(\omega)d\omega$, $\omega\in\R;$

         -a study of conditions necessary and sufficient for
         the symmetric boundary problem described in \S4 to be
         essentially self-adjoint;

         -a generalization  of quantum stochastic calculus to equations
         with nonadapted stochastic differentials extending
         the Ito multiplication table.

\bigskip

\end{document}